\documentclass[twocolumn,notitlepage]{aastex61}

\usepackage{hyperref}
\usepackage{latexsym}
\usepackage{graphicx,epsf,epsfig}
\usepackage{amssymb,amsmath}

\graphicspath{{./figures/}}

\submitjournal{ApJ}

\begin{document}
\title{The arrow of time in the collapse of collisionless self-gravitating systems:\\
  non-validity of the Vlasov-Poisson equation during violent
  relaxation} \author{Leandro {Beraldo e Silva}}
\affiliation{Universidade de S\~ao Paulo, Instituto de Astronomia,
  Geof\'isica e Ci\^encias Atmosf\'ericas, Departamento de Astronomia,
  CEP 05508-090, S\~ao Paulo, SP, Brasil} \email{lbs@usp.br}
\author{Walter {de Siqueira Pedra}}\affiliation{Universidade de S\~ao
  Paulo, Instituto de F\'isica, Departamento de F\'{\i}sica
  Matem\'atica, CP 66318, CEP 05314-970, S\~ao Paulo, SP, Brasil}
\author{Laerte Sodr\'e}\affiliation{Universidade de S\~ao Paulo,
  Instituto de Astronomia, Geof\'isica e Ci\^encias Atmosf\'ericas,
  Departamento de Astronomia, CEP 05508-090, S\~ao Paulo, SP, Brasil}
\author{Eder L. D. Perico}\affiliation{Universidade de S\~ao Paulo,
  Instituto de F\'isica, Departamento de F\'{\i}sica Matem\'atica, CP
  66318, CEP 05314-970, S\~ao Paulo, SP, Brasil} \author{Marcos
  Lima}\affiliation{Universidade de S\~ao Paulo, Instituto de
  F\'isica, Departamento de F\'{\i}sica Matem\'atica, CP 66318, CEP
  05314-970, S\~ao Paulo, SP, Brasil}

\begin{abstract}
  The collapse of a collisionless self-gravitating system, with the
  fast achievement of a quasi-stationary state, is driven by violent
  relaxation, with a typical particle interacting with the
  time-changing collective potential. It is traditionally assumed that
  this evolution is governed by the Vlasov-Poisson equation, in which
  case entropy must be conserved. We run N-body simulations of
  isolated self-gravitating systems, using three simulation codes:
  NBODY-6 (direct summation without softening), NBODY-2 (direct
  summation with softening) and GADGET-2 (tree code with softening),
  for different numbers of particles and initial conditions. At each
  snapshot, we estimate the Shannon entropy of the distribution
  function with three different techniques: Kernel, Nearest Neighbor
  and EnBiD. For all simulation codes and estimators, the entropy
  evolution converges to the same limit as N increases.  During
  violent relaxation, the entropy has a fast increase followed by
  damping oscillations, indicating that violent relaxation must be
  described by a kinetic equation other than the Vlasov-Poisson, even
  for N as large as that of astronomical structures. This indicates
  that violent relaxation cannot be described by a time-reversible
  equation, shedding some light on the so-called ``fundamental paradox
  of stellar dynamics''. The long-term evolution is well described by
  the orbit-averaged Fokker-Planck model, with Coulomb logarithm
  values in the expected range $10-12$. By means of NBODY-2, we also
  study the dependence of the 2-body relaxation time-scale on the
  softening length. The approach presented in the current work can
  potentially provide a general method for testing any kinetic
  equation intended to describe the macroscopic evolution of N-body
  systems.
\end{abstract}
\section{Introduction}
\label{sec:introduction}
The derivation of reduced dynamical descriptions of large systems
composed of many particles is a central issue in Statistical
Mechanics. In such systems, the effect of each individual particle is
weak, but the collective action of the particle ensemble generates a
nontrivial potential field acting on each and every particle. One thus
expects that the state of a system with a large number of identical
particles is reflected in the statistical behavior of one typical
particle in the system. The large scale dynamics is then governed by a
set of autonomous equations describing the evolution of the state of
this typical particle\footnote{The typical particle can also be called
  the test particle, although we prefer the former, meaning that it
  represents the behaviour of the vast majority of the particles.}. In
some cases, the effective dynamics emerges in a mathematically
rigorous fashion through a scaling limit. One important point to be
noted is that the effective macroscopic dynamical equations strongly
depend on the particular scaling or regime which is considered: the
same (large) system appears differently on different scales. For
instance, by considering the same system of Newtonian particles
interacting via some well-behaved inter-particle potential, one can
arrive at the Boltzmann, Landau, or Vlasov equation, as the effective
equation for the ``macroscopic'' dynamics, depending on the time and
space scales, as well as the interaction strength regime considered
\citep{Spohn}.

The Boltzmann equation is based on the assumption that particles
interactions are short range, instantaneous and involve only two
particles at a time (binary collisions). The Vlasov equation is
intended to describe the action of smooth collective effects, without
considering two-body interactions. The Landau equation is intended to
treat binary collisions mediated by the Coulomb interaction. It
represents a long-range limit of the Boltzmann equation, and it is
based on the idea that the relaxation is produced by the cumulative
effects of a large number of weak scatterings (in this way, the Landau
equation is also a kind of Fokker-Planck equation), while the
``field'' particles (the scatterers) are assumed to follow rectilinear
trajectories. The Landau equation can also be seen as a long-time
correction to the Vlasov equation -- see \cite{2013EPJP..128..126C}
for further discussions relating these equations.

In this context, one of the deepest and most debated questions in
Physics since the early development of Statistical Mechanics is the
emergence of the arrow of time in the evolution of macroscopic systems
and how to reconciliate it with the time-reversible microscopic laws,
being them classical, quantum or relativistic
\citep[e.g.][]{ehrenfest1959conceptual, 1993PhT....46i..32L,
  1993PhyA..194....1L, 2001LNP...574...39G}. In other words: how to
explain, starting from the time-reversible equations of motion for the
constituent particles, the irreversibility expressed by the second law
of Thermodynamics for the evolution of the system as a whole?

The first to try to solve this problem was L. Boltzmann at the end of
the nineteenth century, introducing the equation which now bears his
name and the so-called H-theorem, which is intended to prove the
entropy increase from mechanical considerations plus statistical
assumptions \citep{brush1976kind}. The Boltzmann equation is a
particular example of a general class of equations, the transport (or
kinetic) equations, which describe the time evolution of the
distribution function $f(\vec{r},\vec{v},t)$, the probability for a
typical particle to be at position $\vec{r}$ and velocity $\vec{v}$
\citep{Landau_PK}.


The basic format of a transport equation is
\begin{equation}
\label{eq:transp_general}
  \frac{df}{dt} = \Gamma[f],
\end{equation}
where the right-hand side represents the physical process relaxing the
system, i.e. driving it to equilibrium, introducing the arrow of
time. The main hypothesis behind the use of a transport equation of
this form is that the state of the system only depends on its
immediately previous state, having no long-term ``memory'' effects,
i.e. that the evolution is Markovian -- see
\cite{balescu1975equilibrium}.

In the case of a molecular gas, the process responsible for driving
the system to equilibrium is represented by the collisions between
molecules. For this reason, $\Gamma[f]$ is traditionally called the
{\it collisional term}. However, for systems evolving through
non-collisional processes, this name can be misleading: relaxation can
in principle also be produced by collective, collisionless
processes. Henceforth, whenever we refer to the right-hand side of the
transport equation, we call it generically the {\it relaxation term},
which can be associated to collisional or collisionless relaxation
processes.

When deriving the relaxation term, one has to introduce statistical
hypotheses related to the type of interactions between the constituent
particles. In the case of a neutral molecular gas, as mentioned above,
one can assume that the interactions are short-range and that each
interaction is a instantaneous binary collision.

Although this discussion has appeared firstly in the study of
molecular gases, it applies to any system composed of many interacting
particles. In particular, to the process of collapse of
self-gravitating systems and the formation of structures in the
universe. A self-gravitating system is composed of $N$ gravitationally
bound particles moving in the presence of the gravitational potential
created by themselves
\citep{1987degc.book.....S,Binney_2008,Saslaw,2003gmbp.book.....H}. They
range from globular clusters, composed of $N\approx 10^5$ stars, to
galaxies ($N\approx 10^{11}$ stars) and dark matter halos, composed of
a giant number of dark matter particles, whose nature is yet
unknown. The study of the macroscopic evolution of these systems is of
fundamental importance for many reasons, e.g.: they represent the
prototype for any long-range interacting system; understanding their
macroscopic evolution helps us to theoretically model (beyond the
parametrization of the results from numerical simulations) the
quasi-stationary state achieved after the relaxation processes and the
main functions characterizing this state, namely the density profile
and velocity distribution; these functions, besides their intrinsic
importance as dynamical diagnostics, are also key ingredients for
other analyses such as those of dark matter direct and indirect
detection experiments.

The main difference between a self-gravitating system and a neutral
molecular gas is that gravity is a long-range interaction, thus
invalidating all of the assumptions involved in the derivation of the
Boltzmann equation for molecular gases\footnote{In the case of
  plasmas, the particles also interact via long-range forces (Coulomb
  interaction). However, the presence of opposite charges produces an
  screening effect, the Debye shielding, effectively shortening the
  interaction range.}
\citep{1966Phy....32.1376P,Padmanabhan_1990}. On the other hand, it is
still possible to estimate the time-scale for relaxation of a
self-gravitating system due to $2$-body processes (long-range
``collisions''). A general expression is
\citep[see][]{1987degc.book.....S, Binney_2008}
\begin{equation}
  \label{eq:tau_col}
  \tau_{col}=k \frac{N}{\ln \Lambda}\cdot\tau_{cr},
\end{equation}
where $k\approx 0.1$, $\Lambda = b_{max}/b_0$. Here, $b_{max}$ is the
maximum impact parameter of the gravitational scatterings, i.e. it
plays the role of an effective screening length. The parameter $b_0$
is associated to a $90^{\circ}$ scattering angle.  -- see
\S\ref{sec:N_dependence_long} -- and $\tau_{cr}$ is the time scale for
a typical particle to cross the system, the crossing time, which is
also of the same order of the dynamical time scale
$\tau_{dyn}=1/\sqrt{\bar{\rho}G}$, where $\bar{\rho}$ is the mean
density and $G$ is the gravitational constant.

In the case of a globular cluster, the collisional relaxation time
scale is $\tau_{col}\approx 10^9$ yr, shorter than its age
($\approx 10^{10}$ yr) and we conclude that the apparent equilibrium
of these objects has the $2$-body relaxation as a plausible mechanism,
whose relaxation term can be modeled with the Fokker-Planck
approximation, which is based on the weak coupling assumption,
i.e. that the deflection angle produced by each ``collision'' is
small. Also, it neglects any kind of collective relaxation effect --
see \S\ref{sec:Fokker_Planck}.

For galaxies and dark matter halos, given the large number of
particles it is possible to show that the $2$-body relaxation
time-scale is $\tau_{col}\gtrsim 10^{17}$ yr, many orders of
magnitude larger than their ages, and therefore this process is not a
plausible mechanism to explain the apparent (both observationally and
in $N$-body simulations) near equilibrium state that can be achieved
by these systems. They are thus called {\it collisionless
  self-gravitating systems} and are the main focus of this work (we do
not consider any dissipative component such as gas or dust).

The process generally accepted as the driver of a collisionless
self-gravitating system to a quasi-stationary state is the interaction
of the typical particle with the time-changing collective
gravitational potential during the first stages of the collapse of the
system \citep{1962AJ.....67..471K,Henon_1964,LyndenBell_1967}. It is
therefore a collective effect, in contrast to $2$-body relaxation. The
time-scale for that relaxation process, according to some theoretical
arguments \citep{LyndenBell_1967,1990PhyA..169...73K} and to results
from $N$-body simulations, is the dynamical time scale, which is
orders of magnitude smaller than the age of any self-gravitating
system. This process is called {\it violent relaxation}
\citep{LyndenBell_1967, Shu_1978, Madsen_1987, 1987ApJ...316..502S,
  Efthymiopoulos_2007, Bindoni_Secco_2008,2014PhR...535....1L}.
Considering the N-body problem itself, \cite{1986A&A...160..203G} have
derived, based on the Ergodic theory, another relaxation time-scale
associated to collective effects, namely
$\tau \propto N^{1/3}\tau_{cr}$, which is still orders of magnitude
smaller than Eq.~\eqref{eq:tau_col}.

It is interesting to remember that the relaxation process of a general
$N$-body problem is usually related to the presence of stochastic
motions that allow the particles to occupy large regions of phase
space and the system to forget the initial conditions, the so-called
\emph{chaotic mixing} \citep{Merritt_1996_2}. In fact, $N$-body
simulations of galaxy formation have shown evidence of very complex
motions in phase space and the fast achievement (in a dynamical
time-scale) of a quasi-stationary state -- see
\cite{1999PASP..111..129M} and references therein (see also
\cite{2003MNRAS.341..927K} on the role of chaotic mixing in violent
relaxation). This seems to indicate that violent relaxation is a real
relaxation process, in the sense that it drives the system
irreversibly towards the equilibrium state \citep[see
also][]{1990PhyA..169...73K}, even though violent relaxation is known
to be incomplete, ending before the achievement of thermodynamical
equilibrium, which would be described by a Maxwell-Boltzmann
distribution \citep{1999PASP..111..129M, Kandrup_1993,
  Efthymiopoulos_2007}. Consequently, the system keeps some
correlation with the initial conditions. Thus, when we refer to the
equilibrium state generated by violent relaxation, we are actually
considering this incompletely relaxed, quasi-stationary state, which
generally does not correspond to the full thermodynamical equilibrium.

On the other hand, since the $2$-body relaxation of a collisionless
self-gravitating system during the early stages of the collapse is by
definition negligible, it is usually assumed that the relaxation term
associated to violent relaxation is zero. In this case, the system's
evolution is described by Eq.~\eqref{eq:transp_general} with
$\Gamma\left[f\right] = 0$:
\begin{equation}
\label{eq:vlasov}
  \frac{df}{dt} \equiv \frac{\partial f}{\partial t} +
  \vec{v}\cdot\frac{\partial f}{\partial \vec{r}} -
  \frac{\partial \phi}{\partial \vec{r}}\cdot\frac{\partial f}{\partial \vec{v}} = 0,
\end{equation}
where $\phi(\vec{r},t)$ is the gravitational potential associated to
the system as a whole, considered as an external potential for the
typical particle: $d\vec{v}/dt=-\nabla\phi$. This equation is
generically called the Vlasov equation, which can encompass many
different equations, depending on the two-body potential involved. For
the specific problem discussed in this work, namely the evolution of
self-gravitating systems, the two-body potential is Coulombian,
i.e. $\propto 1/r$, and the global potential $\phi$ is
self-consistently related to the distribution function $f$ by means of
the Poisson equation
\begin{equation}
\label{eq:poisson}
  \nabla^2\phi = 4\pi G \int d^3\vec{v}\,f(\vec{r},\vec{v},t).
\end{equation}
Eq.~\eqref{eq:vlasov}, coupled to Eq.~\eqref{eq:poisson}, is then
called the Vlasov-Poisson equation.

Note that, despite the formal similarity of the Vlasov-Poisson
equation with the Liouville equation
\begin{equation}
\label{eq:liouville}
  \frac{df^{(N)}}{dt} = 0,
\end{equation}
where $f^{(N)}(\vec{r}_1,\vec{v}_1,...,\vec{r}_N,\vec{v}_N,t)$ is the
$N$-particle distribution function, in general they represent
different descriptions for the system's
evolution. Eq.~\eqref{eq:liouville}, whose validity is based on
mechanical considerations only, describes the evolution of the total
system and must be valid under very general conditions (for instance,
if the system is subject to external Hamiltonian influences, i.e. if
it is described by a time-dependent Hamiltonian).

On the other hand, Eq.~\eqref{eq:vlasov} refers to the coordinates of
one single typical particle while it is also intended to describe the
evolution of the system as whole, and thus is clearly based on
mechanical plus statistical considerations, what is made explicit in
the construction of the BBGKY hierarchy: as pointed out e.g. by
\cite{Beraldo_Lima_Sodre_Perez_2014}, to reduce the full hierarchy of
$N$-body evolution equations to an (effective) one-body problem one
needs to assume the molecular-chaos hypothesis, i.e. that
$f^{(N)}(\vec{r}_{1},\vec{v}_{1},...,\vec{r}_{N},\vec{v}_{N})$ can be
written as a $N$-fold product of one-particle distribution
functions. Such an assumption is also behind the derivation of other,
more general, effective equations.

Additionally, to derive the Vlasov-Poisson from the $N$-body problem,
it is usually assumed \citep[see][]{NYAS:NYAS28} that the
gravitational $N$-body problem converges to the continuous limit for
$N\rightarrow \infty$, in some sense. Now define a system as being
composed of only one typical particle moving under the influence of
the continuous collective potential, this system being thus described
by a Hamiltonian, which is time-dependent during violent
relaxation. Since Liouville equation is valid even with a
time-dependent Hamiltonian, as stressed above, it should be valid in
this case, and the Vlasov-Poisson equation is interpreted as the
Liouville equation applied to this one-particle system. In this way,
the statistical considerations under the Vlasov-Poisson equation are
apparently erased and the description becomes purely mechanical. In
our opinion, nevertheless, the continuum limit hypothesis only refers
to the mean-field approach to the $N$-body problem, which corresponds
to the use of self-consistent conditions, in the limit of large $N$,
which could depend on many-point correlations. In order to justify the
effectiveness of the corresponding one-body, self-consistent, problem,
the statistical independence of typical particles is still
required. In other words: even if one can consider the Liouville
equation applied to any 1-particle phase-space density $f$, associated
with the potential generated by the large system, it is not clear from
the beginning that such a density $f$ exists for a typical particle
representing the system as a whole, unless some justifiable
statistical assumption is made.

In addition, it has been already shown
\citep[see][]{2000AdSAC..10..229V, 2001PhRvE..64e6209K,
  2002ApJ...580..606H} that the gravitational $N$-body problem
\emph{does not} converge to the continuous limit for
$N\rightarrow\infty$, at least when using Lyapunov exponents as a
diagnostic. Note that being a discrete sample with $N$ bodies (stars
or dark matter particles), as opposed to the continuous limit, is not
a feature of $N$-body simulations only. Most importantly, it is a
feature of Nature. Even in the continuous limit, the description of a
collisionless self-gravitating system's evolution in terms of a
transport equation is not trivial, specially in the presence of
non-integrable potentials generating stochastic orbits -- see
\cite{1982MNRAS.201...15B, NYAS:NYAS28, 1999PASP..111..129M,
  2005NYASA1045....3M} for critical discussions relating the presence
of stochastic orbits, the $N$-body problem and the Vlasov-Poisson
equation.

Furthermore, the main problem in describing the system's evolution
with the Vlasov-Poisson equation is that this equation is time
reversible, while it is intended to describe the irreversible
evolution driven by violent relaxation. This has been already called
the ``fundamental paradox of stellar dynamics''
\citep{1965dss..book.....O, 2006AAT...25..123O}.

In fact, this uncomfortable situation can be seen in several
  works. According to \cite{Madsen_1987}, ``\emph{to reach the
    predicted most probable final state, the system may have to break
    Liouville's theorem} [Vlasov-Poisson
  equation]''. \cite{1987ApJ...316..502S} argues that ``\emph{while
    Liouville's theorem} [Vlasov-Poisson equation] \emph{does apply on
    a microscopic level, it must necessarily be violated on a
    macroscopic level if the concept of violent relaxation is to have
    sensible meaning}''. According to \cite{NYAS:NYAS28}, ``\emph{the
    N-body problem appears to be chaotic on a time scale $∼\tau_{cr}$,
    but the flow associated with the CBE} [Vlasov-Poisson equation]
  \emph{is often integrable or near-integrable in the sense that many
    or all of the characteristics are regular, i.e., non chaotic. So
    what do the (often near-integrable) CBE characteristics have to do
    with the true (chaotic) N-body problem?  (...) The correct answer
    to the question raised above (...) is not completely clear. What
    does, however, seem apparent from the preceding is that, even for
    very large N, true $N$-body trajectories could differ
    significantly from CBE
    characteristics}''. \cite{2002ApJ...580..606H} claim that
  ``\emph{if the rate of growth of small perturbations remains
    substantial even for large N, there would be an important sense in
    which the CBE does not correctly describe the behavior of $N$-body
    systems}''. Finally, \cite{Bindoni_Secco_2008} say that:
  ``\emph{any further relaxation of the system should be therefore
    considered in terms of the coarse-grained phase-space density
    which, as we have seen, would yield results different from the
    predictions based on the initial fine-grained phase-space
    density. This is a worrying aspect of these theories (...). The
    predictions of the theory, based on the fine-grained density, will
    then give a wrong result}''.

  The standard solution to this apparent paradox is to advocate a
  coarse-grain interpretation to the evolution, according to which the
  irreversibility is introduced by our inability to follow the
  phase-space density evolution with absolute precision. In this
  picture, the transport equation, supposedly the Vlasov-Poisson
  equation, refers to the fine-grained distribution function, while
  the irreversible evolution is described by the coarse-grained
  distribution function. Thus, in this standard picture, it is
  difficult to see any relation between the assumed transport equation
  (which refers to the fine-grained distribution function) and the
  physical phenomenon it is intended to describe. Furthermore, in our
  view \citep[see also][and
  \S\ref{sec:final_remarks}]{1965AmJPh..33..391J}, this interpretation
  introduces an undesired subjective element, making the system
  evolution dependent on observations.

  In defense of that standard solution, some authors argue that in the
  fine-grain level (in the continuous limit), the system develops
  phase-space structures that are too fine to be followed by any
  estimator using a finite number of particles, and then in practice
  we always deal with the coarse-grained distribution function -- see
  e.g. \cite{LyndenBell_1967}. Let us stress once more that real
  self-gravitating systems always display a finite number of
  particles, such that one can hardly provide a meaningful
  interpretation to continuous densities having structures finer than
  the typical nearest neighbor (phase-space) distance for the given
  $N$-body system. Globular clusters, for example, which are also
  expected to violently relax in their early evolution, are composed
  of $N\approx 10^6$ stars, which is the number of particles in the
  largest $N$-body simulation used in our analysis. In this sense, any
  coarse graining operating in our results due to the use of a finite
  $N$ is expected to be the same as that operating in real
  self-gravitating systems.


  In this work, we argue that the apparent paradox disappears once we
  abandon the assumption of validity of the Vlasov-Poisson equation
  during violent relaxation. This assumption seems to be due to an
  oversimplifying treatment of the very singular gravitational
  (Coulomb) potential and also to a neglect of the proper statistical
  content of the distribution function regarding the discrete nature
  of the physical system.

  Let us remind that there is no mathematically rigorous proof of the
  validity of the Vlasov-Poisson equation for self-gravitating
  systems. In Appendix \ref{sec:math_vlasov} we give a summary of
  recent mathematical results on the derivation of this equation from
  the $N$-body problem. These suggest that interactions involving
  impact parameters up to scales that are large compared to the mean
  neighboring particle distance $\bar{d}$ could prevent the
  Vlasov-Poisson equation from being the effective macroscopic
  equation governing the evolution of large gravitational systems. As
  discussed below, the numerical results obtained here point in the
  same direction and moreover suggest that violent relaxation does not
  involve scales much smaller than $\bar{d}$.

  Interestingly, studies based on the numerical integration of the
  Vlasov-Poisson equation sometimes obtain results comparable to those
  obtained from $N$-body simulations (although frequently simulating
  situations and scales different from those associated to the violent
  relaxation process), attributing any difference to $N$-body codes
  limitations
  \citep{2013ApJ...762..116Y,2015MNRAS.450.3724C,2016MNRAS.455.1115H}.

  As is well known, if the Vlasov-Poisson equation is valid, then the
  entropy must be conserved
  \citep[see][]{Tremaine_Henon_Lynden_Bell_1986} -- see
  \S\ref{sec:test_vlasov}. In the present work we use $N$-body
  simulations, which are described in \S\ref{sec:simulations}, to
  estimate the entropy of the system at each snapshot, following its
  time evolution. The estimators involved in the entropy estimate are
  presented in \S~\ref{sec:estimators}. In \S\ref{sec:results} we show
  our main results, focusing in the short-term entropy production
  during violent relaxation. Then, in \S~\ref{sec:two_body_relax} we
  analyze the long-term entropy evolution and its description in terms
  of a Fokker-Planck equation. Finally we conclude in
  \S\ref{sec:conclusions}, with further comments in
  \S\ref{sec:final_remarks}.

\section{Testing Vlasov-Poisson equation}
\label{sec:test_vlasov}
Going back to the general form of the transport equation, we have
\begin{equation}
\label{eq:transport_general}
  \frac{df}{dt} \equiv \frac{\partial f}{\partial t} +
  \vec{v}\cdot\frac{\partial f}{\partial \vec{r}} -
  \frac{\partial \phi}{\partial \vec{r}}\cdot\frac{\partial f}{\partial \vec{v}} = \Gamma[f].
\end{equation}

As for any good relaxation process, the relaxation term is responsible
for entropy increase. In fact, following
\cite{Tremaine_Henon_Lynden_Bell_1986}, let
\begin{equation}
  \label{eq:1}
  S = -\int s[f] \,d^3\vec{r}\,d^3\vec{v},
\end{equation}
where $s[f]$ is some functional of the distribution function $f$. For
example, if $s[f]=f\ln f$ then $S$ is the well-known Shannon entropy
associated to the distribution $f$. Accordingly,
\begin{equation}
  \label{eq:S_dot_1}
  \frac{dS}{dt} = -\int \frac{d s}{df}\frac{\partial f}{\partial t} \,d^3\vec{r}\,d^3\vec{v}
\end{equation}
and using the transport equation \eqref{eq:transport_general}:
\begin{equation}
\begin{split}
  \label{eq:S_dot}
  \frac{dS}{dt} &= -\int \frac{d s}{df}\left[\Gamma[f] -
    \vec{v}\cdot\frac{\partial f}{\partial \vec{r}} + \frac{\partial
      \phi}{\partial \vec{r}}\cdot\frac{\partial f}{\partial
      \vec{v}}\right]\,d^3\vec{r}\,d^3\vec{v}\\
  &=-\int \frac{d s}{df}\Gamma[f]\,d^3\vec{r}\,d^3\vec{v} + \\
  & +\int\left[ \vec{v}\cdot\frac{\partial s}{\partial \vec{r}} -
    \frac{\partial \phi}{\partial \vec{r}}\cdot\frac{\partial
      s}{\partial
      \vec{v}}\right]\,d^3\vec{r}\,d^3\vec{v}\\
  &=-\int \frac{d s}{df}\Gamma[f]\,d^3\vec{r}\,d^3\vec{v}.
\end{split}
\end{equation}
In the last passage, we integrate the term
$\vec{v}\cdot\partial s/\partial \vec{r}$ firstly in $d^3\vec{r}$ and
the term
$\partial \phi/\partial \vec{r}\cdot\partial s/\partial \vec{v}$
firstly in $d^3\vec{v}$, using the fact that $s[f]\rightarrow 0$ for
$\vec{r},\vec{v}\rightarrow\infty$.

Thus, if the Vlasov-Poisson equation is valid, i.e. if
$\Gamma[f] = 0$, then $S$, and particularly the Shannon entropy, is
conserved
\citep[see][]{Shu_1978,Tremaine_Henon_Lynden_Bell_1986}. This is to be
expected, since the Vlasov-Poisson equation is time-reversible and
reversible processes keep the entropy constant. On the other hand, if
the quantity defined by Eq.~\eqref{eq:1} is not conserved, this is a
clear evidence for the non-validity of the Vlasov-Poisson equation and
of the emergence of the arrow of time.

The main objective of this work is to test the validity of the
Vlasov-Poisson equation during the violent relaxation of collisionless
self-gravitating systems. We do this using $N$-body simulations to
estimate the entropy of the system at each time, verifying whether it
is conserved. While interesting works on the relaxation of
self-gravitating systems focus on characterizing the mixing properties
of the evolution by means of e.g. calculating Lyapunov exponents or
fundamental frequencies
\citep{Merritt_1996_2,1998ApJ...506..686V,2001PhRvE..64e6209K,2003MNRAS.341..927K},
the key quantity for the relaxation concept, which may be considered
to define it, is the entropy. In this sense, this work go straight to
the \emph{relaxation} concept, with no regards to its explanation in
terms of \emph{mixing}.

In what follows, we will set $s[f]=f\ln f$, i.e. we will only
consider the Shannon entropy
\begin{equation}
  \label{eq:S_def}
  S = -\int f\ln f \,d^3\vec{r}\,d^3\vec{v}.
\end{equation}

\section{$N$-body simulations}
\label{sec:simulations}
We use the code NBODY-6 accelerated with a graphics processing unit
(GPU) -- see \cite{2012MNRAS.424..545N}, which is the result of a long
development since its first version NBODY-1
\citep{1999DDA....30.0601A}. It is a direct summation code and the
integration is based on the scheme of \cite{1973JCoPh..12..389A} using
the forth-order Hermite method \citep{1992PASJ...44..141M}. The code
does not make use of a softening in the Newtonian force law, as
commonly done to avoid close encounters (see below). Instead, it
implements regularization procedures in order to deal with the
possible close encounters, binaries and higher-order objects etc. For
a general discussion about these techniques, see
\cite{2003gnbs.book.....A}.

We use NBODY-6 to simulate an isolated self-gravitating system with
$N$ particles of equal mass $m$, with an initial Maxwellian velocity
distribution with velocity dispersion determined by the initial virial
ratio $Q_0$, which is defined as the ratio of kinetic energy $T$ and
potential energy $W$, $Q = T/|W|$. For our fiducial simulation run, we
use $N=10^5$ particles with a top-hat initial condition, i.e. a
spherically symmetric and spatially homogeneous system, with
$Q_0=0.5$, the value expected at equilibrium.

We also run simulations with different numbers of particles and
different initial conditions: setting $Q_0=0.25$ or $Q_0=0.6$ (the
``cold'' and ``hot'' initial conditions, respectively) for the top-hat
configuration. Additionally we run a simulation with the initial
condition set to the self-consistent Plummer model which is a
stationary state -- see \S\ref{sec:init_cond}. In all simulation runs,
the number of escaping particles, when it occurs, is completely
negligible. The maximum allowed energy relative error was set to
$5\times 10^{-5}$.

Ideally, a $N$-body simulation code intended to simulate collisionless
systems need to run with a (possibly impracticable) large number of
particles in order to suppress $2$-body relaxation, at least for the
time-scales of interest -- see Eq.~\eqref{eq:tau_col}. A widely used
strategy is to introduce the softening length parameter $\varepsilon$,
modifying the Newtonian gravitational potential to the
Plummer-softened one
\begin{equation*}
  \phi(r) = - \, \frac{G \, m}{\sqrt{r^2 + \varepsilon^2}},
\end{equation*}
and thus avoiding close encounters and suppressing the $2$-body
relaxation. In order to study the role of the softening length in the
relaxation process, we also run the simulation code NBODY-2
\citep{2001NewA....6..277A} that, differently from NBODY-6, does make
use of a softening length. This code is also of direct summation type,
but the use of the softening length simplifies the treatment of close
encounters in comparison to NBODY-6. Unfortunately, there is no
parallelized or GPU accelerated version of NBODY-2, what restricts the
possibility of using large numbers of particles.

With the aim of testing the universality of our NBODY-2 and NBODY-6
entropy estimation, we also run the publicly available GADGET-2 code
\citep{2005MNRAS.364.1105S}.  GADGET-2 is a hybrid N-body code that
combines the traditional evolution of self-gravitating collisionless
particles, with the smoothed particle hydrodynamics (SPH) treatment
for collisional gas.  In our case, we run GADGET-2 using the same
initial conditions that were used for NBODY-2 and NBODY-6 simulations,
i. e., an isolated set of self-gravitating collisionless particles,
within a Newtonian space (without Hubble expansion), and without the
presence of any collisional gas.  In addition, we run GADGET-2 using
its tree configuration, i. e., without using the Fourier techniques to
compute long-distance forces.

\begin{figure}
  \raggedright
  \includegraphics[width=9cm]{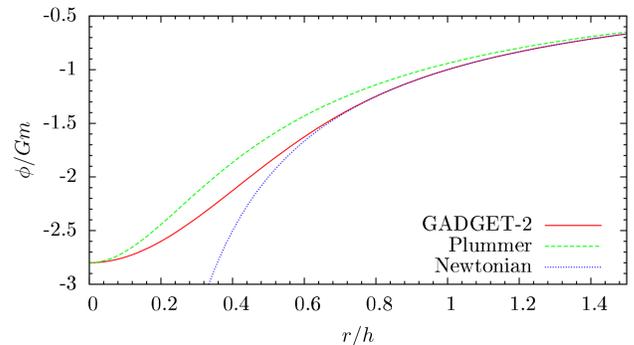}
  \vspace{-0.8cm}
  \caption{Comparison between the spline-softened GADGET-2, the
    Plummer-softened and the Newtonian gravitational potentials of a
    point mass. It was used here $h=1.0$ and $\varepsilon=h/2.8$.}
  \label{fig:softening}
\end{figure}

In order to suppress large-angle scattering during $2$-body
encounters, the GADGET-2 code uses the spline-softened gravitational
potential:
\begin{equation*}
\phi(r)= \frac{G\,m}{h}\,\, W(r/h)\,,
\end{equation*}
where
\begin{equation*}
W(u) =
\begin{cases}
  \cfrac{16}{3}u^2 - \cfrac{48}{5} u^4 +\cfrac{32}{5}u^5-\cfrac{14}{5}\,, & 0\leqslant u <\frac{1}{2}\,,\vspace{0.2cm}\\
  \cfrac{1}{15\,u}+\cfrac{32}{3}u^2-16\,u^3\vspace{0.1cm}\\
  +\cfrac{48}{5}u^4-\cfrac{32}{15}u^5-\cfrac{16}{5}\,, & \frac{1}{2}\leqslant u <1\,,\vspace{0.2cm}\\
  -\cfrac{1}{u}\,, & 1\leqslant u\,,
\end{cases}
\end{equation*}
is based on the \cite{1985A&A...149..135M} SPH modelling, which was
constructed guided by the accuracy, smoothness, and computational
efficiency criteria. This spline-softened potential is equal to the
Newtonian gravitational potential for distances greater than the
softening length $h$, unlike the Plummer-softened potential used by
NBODY-2, which converges slowly to the Newtonian one for long
distances -- see Fig.~\ref{fig:softening}.  In addition, we will refer
to $\varepsilon_{eq}\equiv h/2.8$ as the Plummer-equivalent softening
length for a given $h$ set in GADGET-2. With this choice, the minimum
of the GADGET-2 and the Plummer-softened gravitational potentials have
the same depth at $r=0$, as shown in Fig. \ref{fig:softening}.

All units adopted here are the same as those used internally and
prompted by the simulation code, the H\'enon units
\citep[see][]{Henon_1964}, also called $N$-body units, where the
gravitational constant is $G=1$, total mass $M=1$, and total energy
$E = -1/4$ -- see Appendix \ref{sec:units}. In virial equilibrium,
this corresponds to a virial radius $R_{vir}=1$ and rms velocity
${\sqrt{\langle v^2\rangle}=\sqrt{2}/2}$, and thus the mean crossing
time ${\tau_{cr}=2R_{vir}/\sqrt{\langle v^2\rangle}=2\sqrt{2}}$ -- see
\cite{2001NewA....6..277A} and Appendix \ref{sec:units}. Importantly,
in $N$-body units, positions and velocities are dimensionless
quantities and have values of similar magnitudes.

\section{Entropy estimators}
\label{sec:estimators}
The entropy given by Eq. \eqref{eq:S_def} is estimated using the data
from the $N$-body simulation and translating the integral over phase
space into a sum over the $N$ particles of the system
\citep[see][]{Joe1989,Hall1993,Beirlant1997a}. The heuristics behind
entropy estimators is as follows: having an estimate
$\hat{f}_i=\hat{f}(\vec{r}_i,\vec{v}_i,t)$ for the distribution
function $f$ at the phase space position of each particle $i$, we
estimate the entropy at each time as
\begin{equation}
  \label{eq:S_estimate}
  \hat{S}(t) = -\frac{1}{N}\sum_{i=1}^N \ln\hat{f}_i.
\end{equation}
The meaning and adequacy of this estimator becomes clear when we
interpret Eq. \eqref{eq:S_def} as the phase space average of $\ln f$.
Assuming that the positions $(\vec{r}_{i},\vec{v}_{i})$ of the
particles in phase space are \emph{independently distributed} with
distribution $f$, for particular choices of $\hat{f}_{i}$ (e.g., the
kernel and nearest-neighbors methods discussed below), it has been
shown that this estimator converges in probability\footnote{This means
  that, given any fixed error $\varepsilon >0$, the probability for
  the estimator to make an error larger than $\varepsilon $ for the
  entropy associated to $f$ tends to zero, as $N\rightarrow \infty $.}
to the Shannon entropy given by Eq.~\eqref{eq:S_def} in the limit
$N\rightarrow \infty$ \citep[][]{Joe1989,Hall1993,Beirlant1997a}. We
remark here that the question whether $\hat{f}_{i}$ is a good
estimator for the distribution function $f$ itself, is not explicitly
addressed by \cite{Joe1989,Hall1993,Beirlant1997a} and related works:
instead, the expression in the r.h.s.  of (\ref{eq:S_estimate}) is
directly proven to be a good entropy estimator for certain technically
convenient choices of $\hat{f}_{i}$. Note that assuming the validity
of any dynamical equation for \emph{one} typical particle as the
effective equation for the macroscopic dynamics includes the
assumption that the many-particle correlations do not participate in
the effective dynamics, i.e., different particles typically evolve
independently. Hence, seeing the phase space coordinates of particles
as independent random variables is part of the hypothesis we are
testing, namely, the validity of the Vlasov-Poisson
equation. Fig. \ref{fig:100k_short} shows the early (i.e. in a few
dynamical time-scales) evolution of this estimation, with the
following estimator for the distribution function.

We estimate the entropy of the distribution function $f$ by means of
three different methods: the Kernel method, the Nearest Neighbor
method \citep[see][]{1986desd.book.....S} and the EnBiD method
developed by \cite{2006MNRAS.373.1293S}. This last method has the
advantage that it is fast and metric free, i.e. it does not need to
define distances in 6-D phase-space, for which we need to put
positions and velocities in the same units, which involves some
arbitrary choice -- see Eq.~\eqref{eq:dist_phase_space} and the
comments below. On the other hand, mathematical results showing the
convergence of Eq.~\eqref{eq:S_estimate} to the real entropy when
$\hat{f}_i$ is the EnBiD estimator do not exist, as far as we know.
Also, in contrast to the Kernel method, the convergence of the Nearest
Neighbor is only well-understood at dimension less than 3. Note that
\cite{Beirlant1997a} only mention the derivation of a rate of
convergence for one-dimensional systems, whereas in the case of Kernel
estimators explicit bounds are known at any dimension -- see
\cite{Joe1989,Hall1993}. For these reasons, the analyses in this work
are generally based on the Kernel estimator and in
\S~\ref{sec:neig_vs_kernel} we explain the Nearest Neighbor and EnBiD
methods and show the qualitative agreement between the different
methods and their apparent convergence for increasing N.

The Kernel method \citep[][]{1986desd.book.....S} models the
distribution at a specified point $i$ as a sum of ``bumps'' centered
at each one of all the other particles $j$:
\begin{equation}
  \label{eq:f_kern}
  \hat{f}_i=\hat{f}(\vec{r}_i,\vec{v}_i,t) = \frac{A}{N}\sum_{j=1}^N\frac{1}{h_j^6}K\left( \frac{D_{ij}}{h_j}\right),
\end{equation}
where $D_{ij}$ is the phase-space distance (6-D) between particles $i$
and $j$
\begin{equation}
  \label{eq:dist_phase_space}
  D_{ij} = \sqrt{(\vec{r}_i - \vec{r}_j)^2 + (\vec{v}_i - \vec{v}_j)^2}
\end{equation}
(using the dimensionless coordinates and velocities provided by
H\'enon units -- see \S~\ref{sec:simulations} and Appendix
\ref{sec:units}) and $K\left( \frac{D_{ij}}{h_j}\right)$ is the kernel
function, which determines the shape of the bumps. Note that in
principle the distance estimator in phase-space $D_{ij}$ would involve
coordinates of different units and some metric is necessary to make
them compatible \citep[see][]{2005MNRAS.356..872A,
  2006MNRAS.373.1293S}. One important thing about this metric is that
it should produce coordinates with the same covariance along all
dimensions, which is approximately provided by the use of H\'enon
units.

The parameter $h_j$, here allowed to vary for different particles (the
variable kernel method), is the window width, and it determines the
width of the bumps. This parameter introduces a certain degree of
arbitrariness, analogous to that associated to the bin definitions of
a histogram: it cannot be too small, thus introducing spurious and
noisy substructures, nor too large, what would erase important
information of the distribution. Choosing this parameters optimally
corresponds to improving the convergence rate of the estimator. Here,
we take $h_{j}$ to be the phase space distance $D_{jn}$ from particle
$j$ to its nearest neighbor (see \S\ref{sec:neig_vs_kernel}), which is
a standard choice \citep[see][sec. 2.6]{1986desd.book.....S}. Let us
emphasize that this does not mean that we only consider the
contribution of the nearest neighbor. Instead, the window width just
determines the narrowness of the Kernel function at the position of
particle $j$, and the contributions (tails) of all particles are
important, because the kernels used are heavy-tailed. Note,
additionally, that in the current study the radius of the system is
approximately one. In particular, $D_{jn}$ is typically $N^{-1/6}$. As
in our simulations $N$ ranges between $10^{4}$ and $10^{6}$, in this
case $D_{jn}$ lies between circa $0.1$ and $0.21$.

The normalization constant $A$ is defined by the condition
\begin{equation}
  \label{eq:normal_kernel}
  A = \frac{1}{\int K(x) \,d^6x},
\end{equation}
$x$ being a vector in 6-D, and we use the kernel
\begin{equation}
  \label{eq:kernel}
  K\left( \frac{D_{ij}}{h_j}\right) = \frac{1}{\left(
      D_{ij}/h_j\right)^8 + 1},
\end{equation}
which implies $A = 8/(\sqrt{2}\pi^4)\approx 5.807\times 10^{-2}$.
Observe that entropy estimators of the form (\ref{eq:f_kern}), whose
kernel $K$ has a ``heavy tail'' (i.e., decays slowly in space, without
destroying integrability) are known to have good convergence
properties. See, for instance, \cite{Hall1993}, sec. 3. The particular
(heavy-tail) kernel, Eq. (\ref{eq:kernel}), was chosen because it has
many symmetries in phase space and a simple, explicit normalization
constant $A$.

Let us emphasize that there is no need to advocate any coarse-grain
interpretation to our estimation of the distribution function besides
that present in real systems with finite $N$, since our estimators are
not based on phase-space averages of regions containing large numbers
of particles. Instead, the distribution function at each phase-space
position is estimated directly from the phase-space coordinates of
each particle, and the estimators used have been shown to converge to
the true entropy in the limit $N\rightarrow\infty$
\citep{Beirlant1997a}. Moreover, as shown in
\S~\ref{sec:two_body_relax}, we are able to describe the observed
long-term entropy evolution by means of the Fokker-Planck equation. In
that case also, there is no need to advocate any extra coarse-grain
interpretation.

\section{Results: early evolution and violent relaxation}
\label{sec:results}
Let us remember that for our fiducial simulation run, we use $N=10^5$
particles with a top-hat initial condition, i.e. a spherically
symmetric and spatially homogeneous system, with an initial Maxwellian
velocity distribution with velocity dispersion determined by the
initial virial ratio set to $Q_0=0.5$, the value expected at
equilibrium.

\begin{figure*}[ht!]
  \epsscale{0.85}
  \plotone{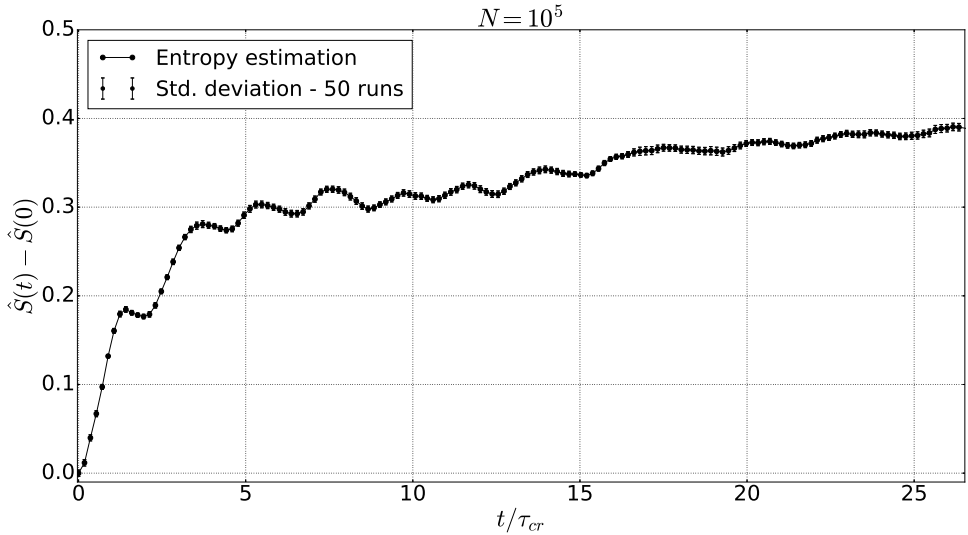}
  \caption{Entropy production estimation $\hat{S}(t) - \hat{S}(0)$,
    Eq. \eqref{eq:S_estimate}, for $N=10^5$ particles starting with a
    homogeneous sphere and a Maxwellian velocity distribution, with
    initial virial ratio $Q_0=0.5$, i.e. with the value expected at
    equilibrium. Time in units of initial mean crossing time
    $\tau_{cr}=2\sqrt{2}$, the expected time-scale for violent
    relaxation. Uncertainties were calculated as the standard
    deviation of 50 runs starting with different seeds for the random
    number generator. The significant entropy increase contrasts with
    the entropy conservation predicted by the Vlasov-Poisson
    equation.}
  \label{fig:100k_short}
\end{figure*}

Fig. \ref{fig:100k_short} shows the initial evolution of the entropy
production $\hat{S}(t) - \hat{S}(0)$, as estimated by
Eq.~\eqref{eq:S_estimate}, with the help of Eqs.~\eqref{eq:f_kern} to
\eqref{eq:kernel}. In this and other plots shown below, time is in
units of initial mean crossing time $\tau_{cr} = 2\sqrt{2}$ -- see
Appendix \ref{sec:units}. The uncertainties were calculated as the
standard deviation of 50 runs starting with different seeds for the
random number generator. We clearly see that the entropy has a
significant increase, accompanied by damping oscillations, in the
dynamical time-scale, which corresponds to the time-scale during which
the violent relaxation process is expected to occur -- see
\S~\ref{sec:introduction}.

These oscillations could, naively, be interpreted as a violation of
the second law of Thermodynamics, which predicts that the entropy must
necessarily increase or be conserved. However, the observed global
entropy increase is in accordance to this. What is apparently violated
is the so-called H theorem, which predicts a monotonic increase of the
H function \emph{if the system's evolution is described by the
  Boltzmann equation}. This apparent violation of the H theorem is,
however, a common feature of any system with an appreciable potential
energy, as argued by \cite{1966Phy....32.1376P,1971PhRvA...4..747J} --
see also \cite{1997JSP....89..735R}. As pointed out by those authors,
these characteristic oscillations are the consequence of the
conversion of kinetic to potential energy and vice-versa, a phenomenon
which is known to occur during the collapse of self-gravitating
systems. In fact, if we assume as a toy-model that the entropy of
these systems has some similarity with that of ideal gases, depending
on the volume $V$ as $\propto \ln V$, the behavior seen in
Fig. \ref{fig:100k_short} can be associated to the system's
macroscopic oscillations with collapse followed by some diffusive
process and the correspondent expansion forming the external
halo. Such entropy oscillations are also well-known in presence of
memory terms in the effective dynamics.

Fig. \ref{fig:100k_short} is the main result of this work. It shows
that during violent relaxation of a $N$-body self-gravitating system
the entropy has a significant increase, while the prediction of
Vlasov-Poisson equation is entropy conservation, what shows that this
equation does not seem to be valid during violent relaxation. In what
follows, we explore other aspects of the problem, namely the role of
the initial conditions, the dependence on the number of particles, the
comparison of different N-body simulators and different distribution
function estimators and finally the role of the softening length in
the codes where it is used. After that we study the long-term
evolution of the entropy, investigating all these aspects in relation
to the two-body relaxation modelled by means of the orbit-averaged
Fokker-Planck equation.

\subsection{Changing initial conditions}
\label{sec:init_cond}
In this section we study what are the consequences of different
initial conditions for the entropy evolution in two different
ways. First, we run the same $N$-body simulations as before, with
$N=10^5$ particles and a top-hat initial spatial distribution with a
Maxwellian velocity distribution, but now changing the initial virial
ratio $Q_0$. Remembering that the expected value at equilibrium is
$Q=0.5$ (the value used in the previous analyses), now we set
$Q_0=0.25$ in one run, which we call ``cold'' initial condition, and
$Q_0=0.6$ in the other (``hot''). And second, in another run, we start
the simulation with a self-consistent Plummer model
\citep[see][]{1974A&A....37..183A}, whose density profile is given by
\begin{equation*}
  \rho(r) = \frac{3M}{4\pi a^3}\frac{1}{\left[1 + \left(r/a\right)^2\right]^{5/2}},
\end{equation*}
where $M$ is total mass and $a$ is a scale factor. In this way, this
simulation run already starts with a steady-state, for which we would
expect no entropy increase.

\begin{figure*}[ht!]
  \epsscale{0.85}
  \plotone{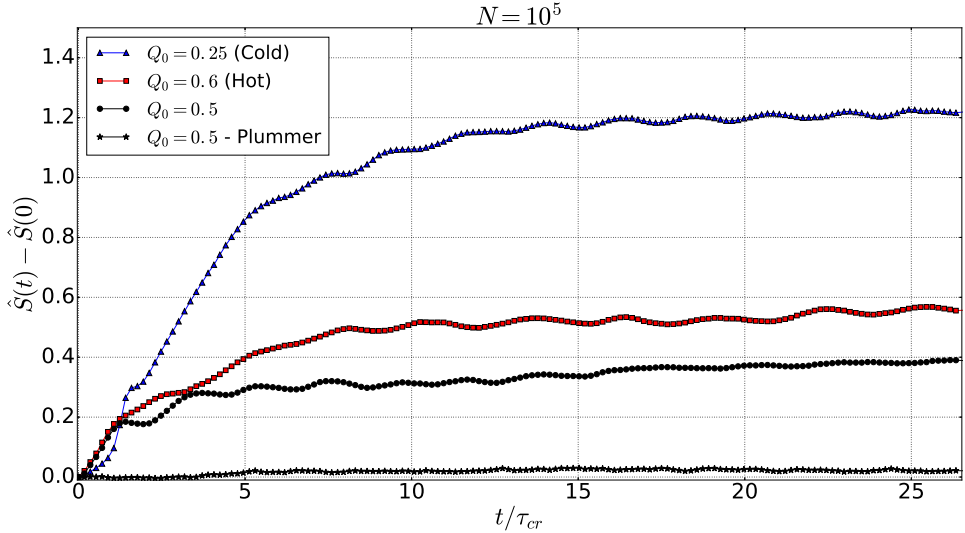}
  \caption{Entropy estimation for $N=10^5$ particles and different
    initial conditions. Blue triangles: initial virial ratio
    $Q_0 = 0.25$ (``cold'' initial condition). Red squares: $Q_0=0.6$
    (``hot'' initial condition). Black dots: the same as in
    Fig.~\ref{fig:100k_short}, the case $Q_0 = 0.5$ (the value
    expected at equilibrium). Black stars: initial self-consistent
    Plummer model, which is a steady-state. Time in units of initial
    mean crossing time $\tau_{cr}=2\sqrt{2}$, the time-scale expected
    for violent relaxation. All these curves are consistent with the
    expectation that the farther from the equilibrium, the larger the
    entropy production.}
  \label{fig:100k_cold_hot_plum_short}
\end{figure*}

Fig. \ref{fig:100k_cold_hot_plum_short} shows the early entropy
evolution for these configurations. We see that all three curves with
a top-hat initial density profile and varying $Q_0$ show the same
qualitative behaviour: a high entropy increase followed by damping
oscillations, in the dynamical time-scale. On the other hand, this
early entropy increase is significantly smaller (virtually negligible)
for the initial Plummer model. In
Fig. \ref{fig:100k_cold_hot_plum_short}, time is again in units of the
initial mean crossing time. Note that this quantity depends on $Q$
(see Appendix \ref{sec:units}), which also changes in time,
oscillating around and converging to the value expected at
equilibrium, $Q=0.5$. For $Q_0=0.25$ and $Q_0=0.6$ the initial mean
crossing time would be $\tau_{cr}\approx 7.35$ and
$\tau_{cr}\approx 1.85$, respectively. However, in order to avoid
confusion, we normalize all curves by the same constant value
$\tau_{cr}(Q=0.5)=2\sqrt{2}$. If we had normalized by the respective
$\tau_{cr}$ values, the blue (cold) curve would be slightly compressed
to the left and the red (hot) curve would be slightly stretched out to
the right.

The most striking feature of this plot is the fact that when we start
with initial conditions farther from equilibrium the entropy increase
is higher, as we would expect based on the second law of
Thermodynamics. For the simulation starting with the self-consistent
Plummer model, the entropy production is almost zero. Therefore, the
large entropy increase observed in the other curves probably cannot be
attributed to any artificial numerical effect. Instead, as in the
general idea of the entropy increase in any macroscopic system, it is
the consequence of the choice of a particular initial state, very
unlikely in comparison to the near-equilibrium state.

\subsection{Dependence on the number of particles}
\label{sec:N_dependence}
The distribution function estimation is expected to represent the true
distribution function and, for significantly large $N$, the
distribution function of a collisionless system.  One concern then is
whether the number of particles used is enough for achieving
convergence to this limit.

\begin{figure*}[ht!]
  \epsscale{0.85}
  \plotone{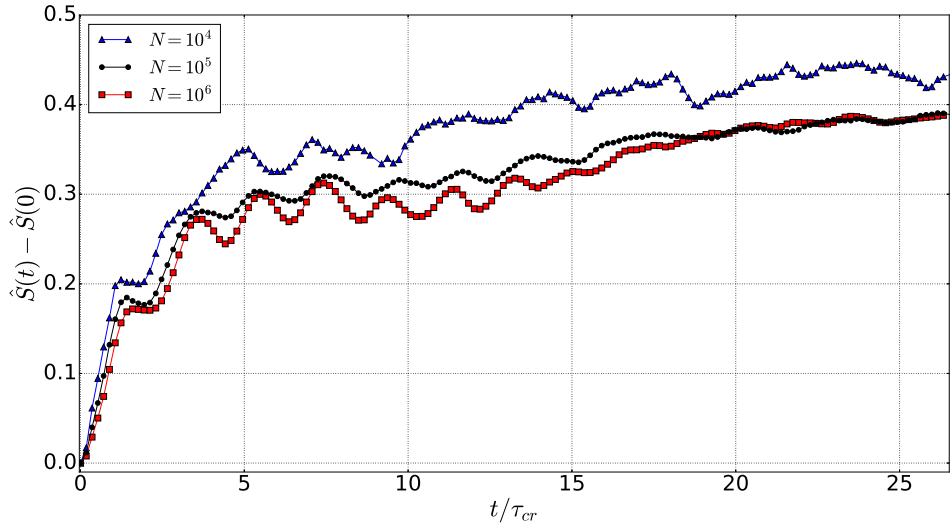}
  \caption{Entropy estimation for different numbers of particles as a
    function of time. Blue triangles for $N=10^4$, black dots for
    $N=10^5$ and red squares for $N=10^6$. The similarity between the
    curves for $N=10^5$ and $N=10^6$ suggests the convergence for
    significantly large $N$.}
  \label{fig:10k_100k_1m_short}
\end{figure*}

Fig. \ref{fig:10k_100k_1m_short} shows the early entropy evolution for
different numbers of particles (blue triangles for $N=10^4$, black
dots for $N=10^5$ and red squares for $N=10^6$). We see that the three
curves have the same qualitative behaviour: a fast entropy increase,
followed by damping oscillations. For $N=10^4$, we still have
significant noise deforming the oscillatory pattern, while for
$N=10^5$ and $N=10^6$ the curves are smoother. Also, the curves for
$N=10^5$ and $N=10^6$ are very similar and achieve the same value at
$t\approx 75$, what seems to indicate the convergence for this early
evolution. This convergence is also important because it indicates
that the high entropy increase during violent relaxation is not due to
two-body relaxation associated to the use of a finite number of
particles.

\subsection{Comparison of distribution function estimators}
\label{sec:neig_vs_kernel}
The second method we use to estimate the distribution function
$f(\vec{r},\vec{v},t)$ is the Nearest Neighbor method
\citep[see][]{1986desd.book.....S}, which is based on the following
idea: sitting on particle $i$, we use Eq.~\eqref{eq:dist_phase_space}
to calculate the (6-D) phase space distance $D_{in}$ to particle $n$,
its nearest neighbor, i.e. the smallest of the distances to all the
other particles $j$, Eq.~\eqref{eq:dist_phase_space}.

This distance is used to define a hyper-sphere of volume
$\propto D_{in}^6$ centered on particle $i$. The estimation of the
distribution function at the phase-space position of particle $i$ is
then
\begin{equation}
  \label{eq:f_neig}
  \hat{f}(\vec{r}_i,\vec{v}_i,t) = \frac{1}{D_{in}^6},
\end{equation}
i.e., it is the number of particles inside the sphere divided by its
volume. In principle, this estimation must be normalized, but we use
it just to estimate the entropy production $S(t) - S(0)$, for which
additive constants, i.e. multiplicative factors in $f$, cancel
out. Note that, differently from the Kernel method,
Eq.~\eqref{eq:f_kern}, the Nearest Neighbor estimation does not
involve a sum over particles. Instead, it only considers the distance
to the nearest neighbor, thus being more prone, at least in principle,
to larger Poisson errors.

\begin{figure*}[ht!]
\epsscale{0.85}
\plotone{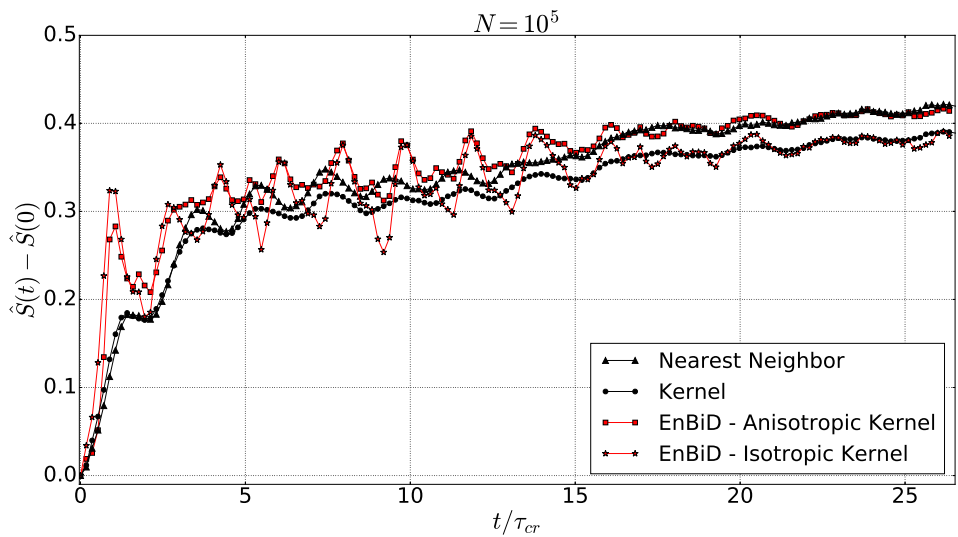}
\caption{Entropy estimation for $N=10^5$ with different methods. Black
  dots (triangles) show the entropy estimation obtained with Kernel
  (Nearest Neighbor) method, Eqs.~\eqref{eq:f_kern} and
  \eqref{eq:f_neig}. Red squares (stars) represent the entropy
  evolution obtained with EnBiD method with anisotropic (isotropic)
  kernel smoothing. Despite some differences, mainly the higher
  initial entropy production obtained with EnBiD in comparison to the
  other methods, the overall behavior of the estimation does not
  change for different estimators: entropy increases accompanied by
  damping oscillations.}
\label{fig:neig_vs_kern_vs_EnBiD_100k_short}
\end{figure*}

\begin{figure*}[ht!]
\epsscale{0.85}
\plotone{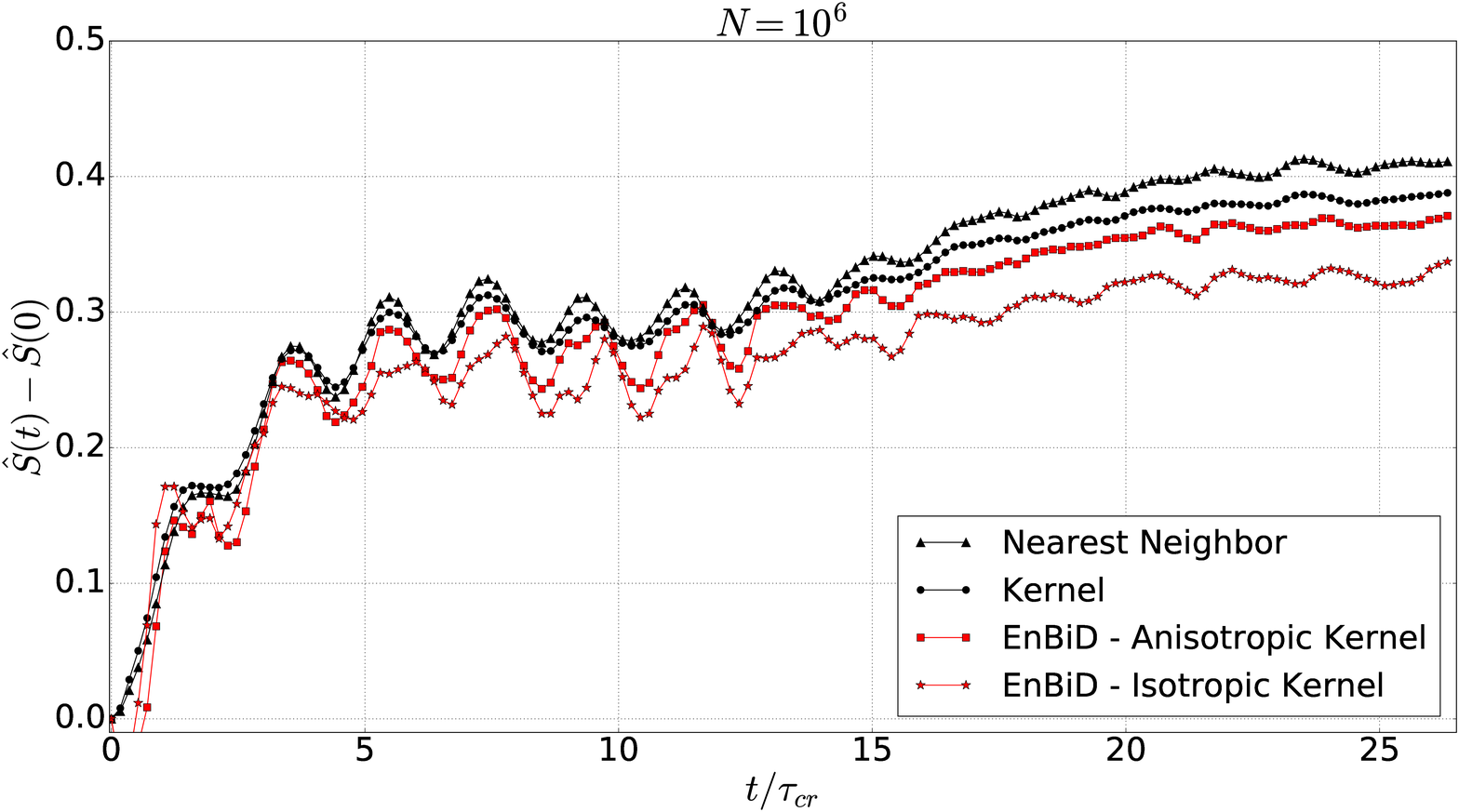}
\caption{Same as Fig.~\ref{fig:neig_vs_kern_vs_EnBiD_100k_short} but
  now for $N=10^6$ particles. The larger initial entropy production
  obtained with EnBiD in comparison to the other methods is not
  present anymore and the oscillation amplitudes of all methods are
  closer. Qualitative changes of EnBiD from $N=10^5$ to $N=10^6$ are
  larger than for the other methods, suggesting a faster convergence
  of Nearest Neighbor and Kernel methods. The similarity of all curves
  also suggests convergence for significantly large $N$.}
\label{fig:neig_vs_kern_vs_EnBiD_1m_short}
\end{figure*}

Figs. \ref{fig:neig_vs_kern_vs_EnBiD_100k_short} and
\ref{fig:neig_vs_kern_vs_EnBiD_1m_short} compare the early entropy
evolution obtained with the different estimator methods, for $N=10^5$
and $N=10^6$ respectively. Results obtained with Nearest Neighbor
method are represented by black triangles. We see that for $N=10^5$,
this method is already very close to the Kernel method estimation
(black dots), although showing larger amplitude oscillations. For
$N=10^6$, Fig.~\ref{fig:neig_vs_kern_vs_EnBiD_1m_short}, the
differences are smaller and the oscillation amplitudes are very close
to those obtained with the Kernel method.

The third distribution function estimator method we use is EnBiD
\citep{2006MNRAS.373.1293S} which is based on the general idea of
binary space partitioning tree, and consists of three steps: first the
phase space is tessellated into mutually disjoint hyper-cubes
containing one particle each. Here, the phase-space density could
already be estimated as $1/V_i$, where $V_i$ is the volume of
hyper-cube $i$, similarly to the Nearest Neighbor method. Then,
boundary corrections are applied to consider the arbitrary shape of
the volume with the data. Finally, in order to reduce noise, that
first density estimation is smoothed by a number of different
techniques, mainly the Kernel method. Also, different kernels can be
chosen. In this smoothing procedure, EnBiD sums the contribution of a
fixed number of neighbors. Note that this is different from what we do
in the Kernel estimator, where we sum over all systems's particles --
see \S~\ref{sec:estimators}. As a generalization, EnBiD also allows
the use of Anisotropic Kernels, what can improve the estimation in
regions with large density variations, such as the borders of the
system -- see \cite{2009MNRAS.393..703M}.

EnBiD has been shown to accurately recover the distribution function
used as input -- \cite{2006MNRAS.373.1293S, 2009MNRAS.393..703M}. Here
we use EnBiD method with the Epanechnikov Kernel smoothing, taking
into account the contribution of 25 neighbors. For comparison, we also
use EnBiD with the anisotropic kernel option. In
Figs. \ref{fig:neig_vs_kern_vs_EnBiD_100k_short} and
\ref{fig:neig_vs_kern_vs_EnBiD_1m_short}, the EnBiD estimates are
represented by red squares and stars, for the Anisotropic and
Isotropic kernels, respectively. For $N=10^5$, we see important
deviations in respect to the other two methods, mainly in the initial
entropy increase and in the oscillation amplitudes, although the
differences being smaller for the Anisotropic version, which we
consider to be more accurate. For $N=10^6$, the EnBiD estimators,
mainly with the Anisotropic kernel, are very close to what is obtained
with the Nearest Neighbor and Kernel methods: that high discrepancy in
the initial entropy production is not present and the oscillation
amplitudes are smaller and similar to those of the other methods. It
is interesting to note that the qualitative changes in EnBiD estimate
from $N=10^5$ to $N=10^6$ are larger than that of the Nearest Neighbor
and Kernel methods, suggesting a faster convergence of the latters.

The important point here is that all the different methods show the
same qualitative behavior for the entropy evolution: a fast monotonic
increase up to the dynamical time-scale, followed by an increase with
damping oscillations up to, at least, $t/\tau_{cr}\approx 25$. The
similarity and smoothness of the curves obtained with different
methods indicate that the features observed really represent the
behaviour of the entropy during the evolution of the simulated system,
and are not spurious effects associated to a particular method. It is
interesting to note the similarity of all the methods for
$N=10^6$. Both the Kernel and the Nearest Neighbor estimators are
known to converge, independently of each other, to the same value
(namely, the entropy of the considered distribution) for sufficiently
large $N$, and the same is expected for EnBiD. Thus, this agreement
suggests that this convergence has already been achieved with $N=10^6$
particles.

\subsection{Comparison of N-body simulators and the role of the
  softening length}
\label{sec:eps}
As mentioned in \S~\ref{sec:simulations}, we also run simulations with
the code NBODY-2, which makes use of a softening length in order to
avoid close encounters and suppress $2$-body relaxation. Varying that
parameter allows us to study the characteristic scales for different
relaxation processes. Some particularly important scales in this
respect are the system's size $R\approx 1$ and the mean neighboring
particle distance $\bar{d}=R/N^{1/3}$.

\begin{figure*}[ht!]
  \plotone{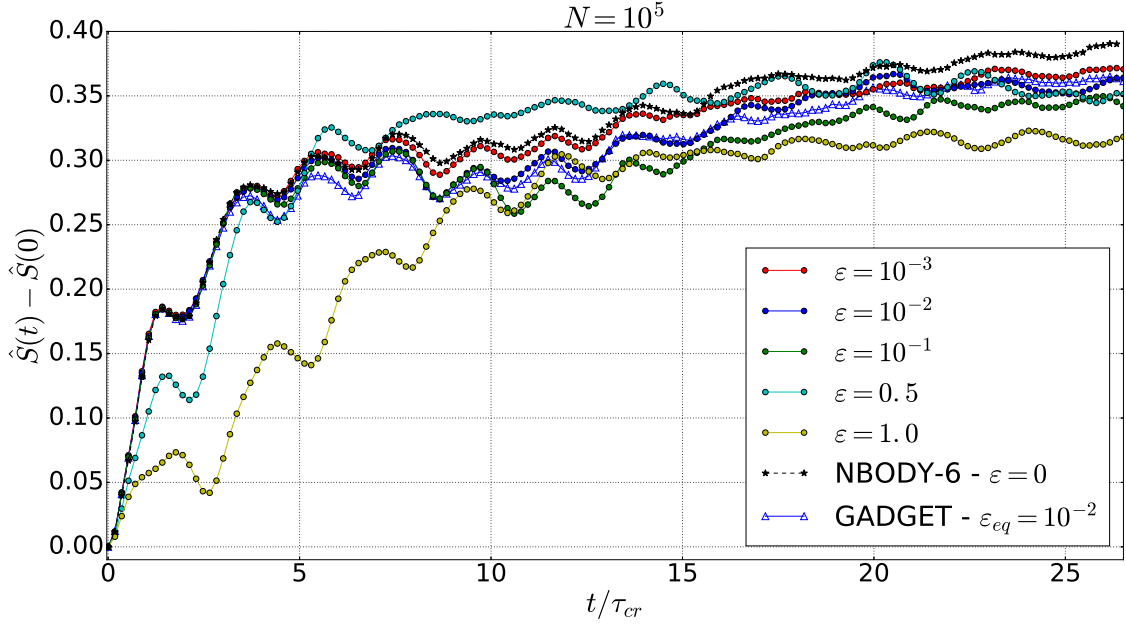}
  \caption{Entropy estimation (Kernel method) for $N=10^5$ particles
    obtained with NBODY-2 for different values of softening length
    $\varepsilon$. Entropy production is suppressed for $\varepsilon$
    values considerably larger than
    $\bar{d}\approx R/N^{1/3}\approx 0.02$.  Black stars represent the
    entropy obtained with NBODY-6 (same as in
    Fig.~\ref{fig:100k_short}), i.e. with $\varepsilon=0$. Also shown
    the entropy obtained with GADGET-2 (blue open triangles) for a
    softening length $h=2.8\times 10^{-2}$, i.e. with a
    Plummer-equivalent $\varepsilon_{eq}=10^{-2}$, whose evolution is
    very similar to that of NBODY-2 with same $\varepsilon$ (blue
    dots).}
  \label{fig:100k_eps_short}
\end{figure*}

Fig.~\ref{fig:100k_eps_short} shows the early evolution of the entropy
for a simulation with $N=10^5$ and the fiducial initial conditions,
i.e. a top-hat with Maxwell velocity distribution and initial virial
ratio $Q_0=0.5$, for different values of the NBODY-2 softening length
$\varepsilon$. Black stars represent the entropy evolution obtained
with NBODY-6, i.e. without any softening ($\varepsilon=0$). We firstly
note the similarity of this curve with the entropy evolution obtained
with NBODY-2 for the smallest softening length used
$\varepsilon=10^{-3}$ (red dots).

It is possible to see that the initial entropy increase is essentially
the same for all runs with $\varepsilon \leq 0.1$, suggesting that
violent relaxation do not involve scales much smaller than the mean
neighboring particle distance $\bar{d}\approx R/N^{1/3}\approx 0.02$.
On the other hand, for $\varepsilon=0.5$ and $\varepsilon=1.0$, we
observe an important suppression in the early entropy production,
suggesting that the violent relaxation process involves scales larger
than $\bar{d}$. We notice that this is in accordance with recent
mathematical results on effective dynamical equations for large
Newtonian systems, i.e. that interactions involving scales around
$\bar{d}$ could prevent Vlasov-Poisson equation to be valid -- see
Appendix~\ref{sec:math_vlasov}.

The delayed entropy increase for these large values of $\varepsilon$
can be interpreted as follows: with such high values of $\varepsilon$,
the system starts evolving almost as if it was composed of
non-interacting particles, for which we would expect the entropy to be
constant or to evolve slowly due to phase-mixing finite-$N$ coarse
graining. However, once the system expands, the typical distance
between particles increases, ``turning on'' the interactions between
particles with distances larger than $\varepsilon$, thus increasing
the entropy production rate.

We also simulate the evolution of one halo with the tree code
GADGET-2, starting with exactly the same initial conditions as those
of the previous simulations, and with a softening length
$h=2.8\times 10^{-2}$, i.e. with a Plummer-equivalent softening length
$\varepsilon_{eq}=10^{-2}$ -- see \S \ref{sec:simulations}. The
entropy evolution is shown as the blue open triangles in
Fig. \ref{fig:100k_eps_short} and we see that it is very similar to
that obtained with NBODY-2 and the same value of the softening length
(blue dots). Since these codes are based on very different integration
techniques, this weakens the possibility of the entropy evolution
observed being due to an artificial numerical relaxation, unless it is
present in exactly the same way in both codes.

\section{Long-term evolution and two-body relaxation}
\label{sec:two_body_relax} Taking into account the collisional
relaxation time scale, Eq.~\eqref{eq:tau_col}, in the limit
$N\rightarrow\infty$ the system becomes collisionless, i.e. the
$2$-body relaxation is expected to be negligible in this limit. The
study of this limit is the main goal of this work, as discussed in the
previous section. However, in N-body simulations we necessarily deal
with a limited number of particles, and some $2$-body relaxation is
always present. In this section, we model the effects of $2$-body
relaxation by means of the orbit-averaged Fokker-Planck equation,
assuming that the potential is static and that the distribution
function is a function of energy only. Thus, this approach applies to
the long-term evolution, after violent relaxation. The terms involved
in the Fokker-Planck relaxation term are estimated with the Agama
library \citep{Vasiliev_2017}, as explained below.

\subsection{Entropy production of a general process}

Given the entropy definition Eq. \eqref{eq:S_def}, we have, according
to Eq. \eqref{eq:S_dot},
\begin{equation}
  \label{eq:18}
  \frac{dS}{dt} = -\int (1 + \ln f)\Gamma[f]\,d^3\vec{r}\,d^3\vec{v},
\end{equation}
which we can estimate as
\begin{equation}
  \label{eq:est_S_dot}
  \widehat{\frac{dS}{dt}} = -\frac{1}{N}\sum_{i=1}^N
  \frac{(1 + \ln \hat{f}_i)}{\hat{f}_i}\Gamma[\hat{f}_i].
\end{equation}
Thus, the theoretical prediction for the entropy produced by any model
through the relaxation term $\Gamma[f]$ can be estimated with the
simulation data, then integrating $\widehat{dS/dt}$ as
\begin{equation}
  \label{eq:num_int_S}
  \hat{S}(t+\Delta t) = \hat{S}(t) + a\cdot\widehat{\frac{dS}{dt}}(t)\Delta t
\end{equation}
and fitting to the simulation data with 2 free parameters, $S_0=S(0)$
and $a$. This can be compared to the entropy production obtained with
the same data, Eq.~\eqref{eq:S_estimate}, configuring a general method
to test any theoretical transport equation, in particular the
Fokker-Planck equation.

\subsection{Fokker-Planck relaxation term}
\label{sec:Fokker_Planck}
The Fokker-Planck relaxation term is used below to estimate the
$2$-body relaxation contribution to entropy production. This model
considers the relaxation of the system as a result of cumulative
effects of many $2$-body weak encounters, with the energy change and
deflection angle in each encounter being small, neglecting any
possible collective relaxation effect, i.e. assuming that encounters
are independent on each other. Following standard procedures
\citep{1987degc.book.....S, Binney_2008, 2003gmbp.book.....H}, we only
consider diffusion in the velocity field. In the local approximation,
the diffusion coefficients are considered to depend only on the
velocity of the test particle, but not on its position, which is valid
for systems with a constant potential. For more realistic systems, in
which the potential does have a spatial dependence, we resort to the
orbit-averaged Fokker-Planck equation, in which case the diffusion
coefficients are averaged over the volume $V$ accessible to the test
particle. For a spherically symmetric static potential and assuming
that the distribution function depends only on energy, the
orbit-averaged Fokker-Planck relaxation term is given by
\begin{multline}
  \label{eq:FP_general}
  \Gamma[f]_{FP} = \frac{1}{g(E)} \left\{-\frac{d}{d
      E}\left[f(E)g(E)\langle
      \Delta E\rangle_V\right] +\right.\\
  \left. +\frac{1}{2}\frac{d^2}{dE^2}\left[f(E)g(E)\langle (\Delta
      E)^2 \rangle_V\right]\right\},
\end{multline}
where $g(E)$ is the density of states:
\begin{equation}
  \label{eq:g}
  g(E) = 16\pi^2 \int_0^{r_m(E)}dr r^2 v,
\end{equation}
with $v = \sqrt{2[E-\phi(r)]}$. The diffusion coefficients
$\langle \Delta E\rangle$ and $\langle (\Delta E)^2\rangle$ are given
by
\begin{equation}
  \label{eq:Delta_E}
  \langle \Delta E\rangle = \gamma (I_0 - I_{1/2}),
\end{equation}
\begin{equation}
  \label{eq:Delta_E2}
  \langle (\Delta E)^2\rangle = \frac{2}{3}\gamma v^2(I_0 + I_{3/2}),
\end{equation}
where $\gamma = 16\pi^2 G^2 m \ln\Lambda$ and
\begin{equation}
  \label{eq:I0}
  I_0(E) = \int_E^0 f(E')dE' = \int_v^\infty f(r,v')v'dv',
\end{equation}
\begin{equation}
  \label{eq:In2}
\begin{split}
  I_{n/2}(E,r) &= \int_{\phi(r)}^E \left(\frac{E' - \phi(r')}{E -
      \phi(r)}\right)^{n/2}f(E')dE' \\
 & = v\int_0^v\left(\frac{v'}{v}\right)^{n+1} f(r,v')dv'.
\end{split}
\end{equation}
Finally, the orbit-average operation is defined as
\begin{equation}
  \label{eq:20}
  \langle ... \rangle_V = \frac{16\pi^2}{g(E)}\int_0^{r_m(E)}dr
  r^2v \langle ... \rangle
\end{equation}
-- see \cite{2015MNRAS.446.3150V} for similar expressions. After some
algebra, we have
\begin{multline}
  \label{eq:FP}
  \Gamma[f]_{FP} = \gamma\Bigg\{\Big[ \langle I_0(E)\rangle_V +
      \langle I_{1/2}(E,r)\rangle_V\Big]\frac{df}{dE}+ \Bigg.\\
+ \frac{1}{3}\Big[ \langle v^2I_0(E)\rangle_V +
      \langle v^2I_{3/2}(E,r)\rangle_V \Big]\frac{d^2f}{dE^2} + f^2(E) \Bigg. \Bigg\}
\end{multline}

We estimate all quantities involved in the orbit-averaged
Fokker-Planck relaxation term using the Agama library
\citep{Vasiliev_2017}, which works in the following steps: for the
$N-$body sample at a given snapshot, the library determines a smooth
global potential $\phi(r)$ and the density of states $g(E)$,
Eq.~\eqref{eq:g}. Given the sample energy distribution $N(E)=f(E)g(E)$
the code calculates smooth functions $f(E)$, $df/dE$ and
$d^2f/dE^2$. The Agama library also provides accurate estimates for
the orbit-average of the integrals involved in the calculation of the
diffusion coefficients, Eqs.~\eqref{eq:Delta_E}-\eqref{eq:In2}.

Substituting these expressions into Eq. \eqref{eq:FP}, we
estimate at each snapshot the contribution of the Fokker-Planck
relaxation term to the entropy increase,
Eq.~\eqref{eq:est_S_dot}. This is then integrated with
Eq. \eqref{eq:num_int_S} and fit to the simulation data, with the free
parameter $a$ being associated to the value of the Coulomb logarithm
$\ln\Lambda$.

Note that, in principle, it is possible to estimate all quantities
involved in Eq.~\eqref{eq:FP} with the same kind of estimators used
here for the entropy. For example, it is possible to estimate the
orbit-averaged diffusion coefficients and, given the estimator for
$f(r,v)$, Eq.~\eqref{eq:f_kern}, to calculate
$\partial \hat{f}/\partial v$ and $\partial^2 \hat{f}/\partial v^2$.
However, when trying this we observed that, although our estimates for
the diffusion coefficients seem very accurate, being essentially
identical to what we get with the Agama library, our estimates for the
derivatives of $f$ are noisy, producing unsatisfactory results. Note
also that, even though we are able to get expressions for terms like
$\partial \hat{f}/\partial v$, i.e. for the derivative of the
estimator, in reality what we need are terms like
$\widehat{\partial f/\partial v}$, i.e. the estimator of the
derivative. In other words, it is not clear if the estimators commute
with the derivatives, and a detailed study of these properties is out
of the scope of this work. In future works, it would be interesting to
better understand these operators and how to accurately estimate these
derivatives, opening the possibility of testing any theoretical
transport equation.

\subsection{Main results on two-body relaxation}
\begin{figure*}[ht!]
  \epsscale{0.85}
  \plotone{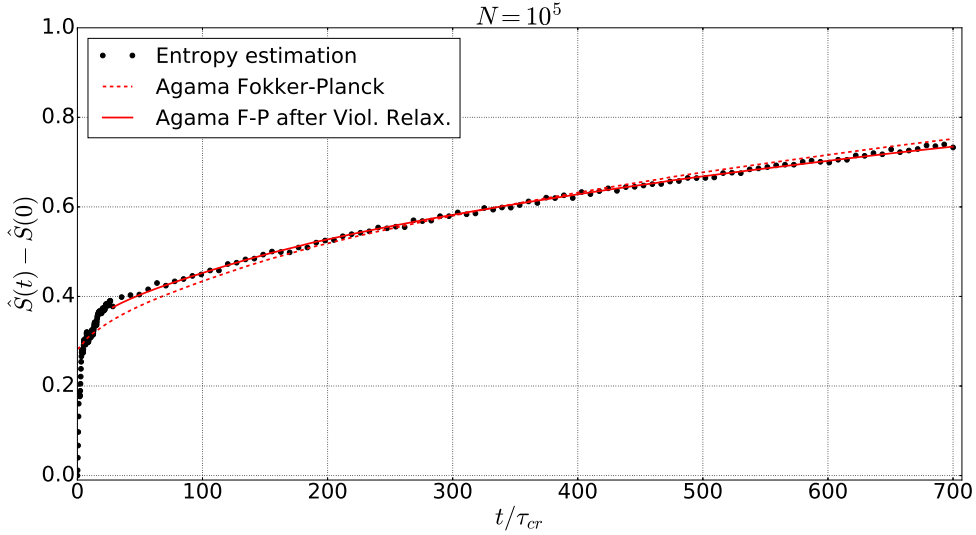}
  \caption{Entropy estimation for $N=10^5$ particles (dots) and
    Fokker-Planck (F-P) prediction estimated by Agama (curves). Note
    the difference in the range of the x-axis compared to
    Fig.~\ref{fig:100k_short}. Dashed line: fit of F-P equation,
    Eqs. \eqref{eq:FP} and \eqref{eq:est_S_dot}, taking into account
    all the data points. The high entropy production during violent
    relaxation at early stages cannot be described by $2$-body
    relaxation as modeled by F-P equation. Solid line: fit of the same
    model, but taking into account just data points for
    $t/\tau_{cr} > 26.5$. The model can fit the entropy evolution in
    this regime.}
  \label{fig:fit_100k}
\end{figure*}

Fig. \ref{fig:fit_100k} shows the entropy production estimation
${\hat{S}(t) - \hat{S}(0)}$ in the long-term evolution ($t/\tau_{cr}$
up to $700$) for the fiducial simulation. After the fast increase at
early times, as shown in Fig. \ref{fig:100k_short}, the entropy growth
becomes slower and almost linear. The dashed curve shows the fit of
the Fokker-Planck model when we consider all the data points.  Since
in the early evolution the hypotheses behind the Fokker-Planck model,
namely a static potential and $f =f(E)$ are not valid, it is not
surprising that the model cannot describe the data in this regime. On
the other hand, if we restrict the fit to later times, say for
$t/\tau_{cr}\gtrsim 26.5$, we see an excellent agreement with data, as
shown by the solid line. Thus this long-term evolution can clearly be
attributed to $2$-body relaxation. Note that the smooth, almost linear
behaviour of the Fokker-Planck prediction is not put by hand, but it
is a consequence of the non-trivial combination of several terms in
Eq. \eqref{eq:FP}.


\subsection{Changing initial conditions}
\begin{figure*}[ht!]
  \epsscale{0.85}
  \plotone{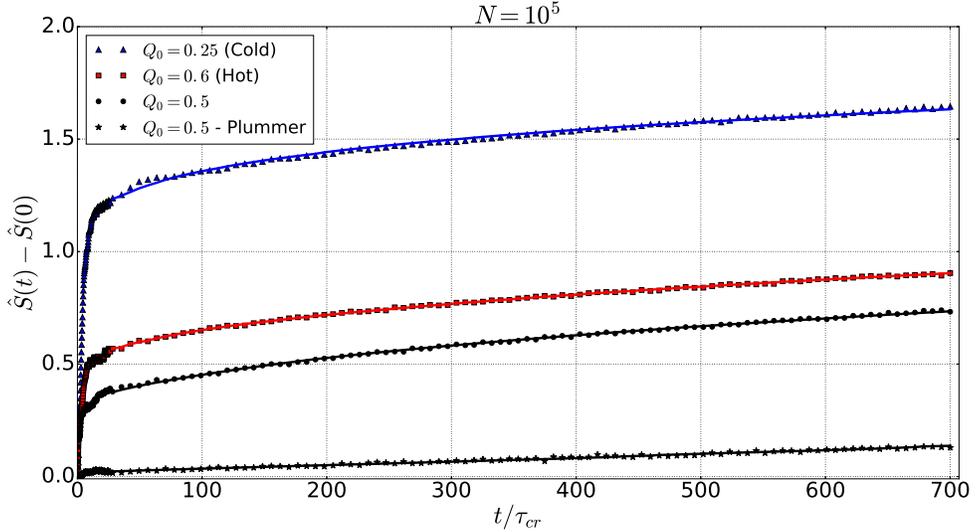}
  \caption{Long-term entropy evolution for different initial
    conditions: top-hat with $Q_0=0.25$ (blue triangles), $Q_0=0.5$
    (black dots) and $Q_0=0.6$ (red squares); and also an initial
    Plummer model (black stars) -- see
    Fig. \ref{fig:100k_cold_hot_plum_short}. The orbit-averaged
    Fokker-Planck equation (curves) can describe the long-term entropy
    evolution (for $t/\tau_{cr}\gtrsim 26.5$), but not the early
    evolution driven by violent relaxation. Two-body relaxation rate
    seems to be insensitive to initial conditions.}
  \label{fig:100k_cold_hot_plum_long}
\end{figure*}

Fig. \ref{fig:100k_cold_hot_plum_long} shows the long-term entropy
evolution for the simulations with different initial conditions
discussed in \S~\ref{sec:init_cond}, namely: top-hat with $Q_0=0.25$
(blue triangles), $Q_0=0.5$ (black dots) and $Q_0=0.6$ (red squares);
and also starting with a Plummer model (black stars). As in the
fiducial case, the Fokker-Planck model (continuous lines) can
reproduce the almost linear entropy increase for
$t/\tau_{cr} \gtrsim 26.5$, but not for early times. While the entropy
increase at early times depends crucially on the initial conditions,
it is basically the same for the long-term evolution (same slope of
$\hat{S}(t)$). In other words, collisional relaxation seems to depend
just on $N$, but not on the initial conditions, while violent
relaxation strongly depends on the initial conditions. Here again we
have time normalized by $\tau_{cr}=2\sqrt{2}$ in all curves
(normalizing by the respective initial values of $\tau_{cr}$ would
give the false impression that the long-term relaxation rate, i.e. the
slope of $\hat{S}(t)$, is different for the different initial
conditions.)

\subsection{Comparison of distribution function estimators} 

\begin{figure*}[ht!]
  \epsscale{0.85}
  \plotone{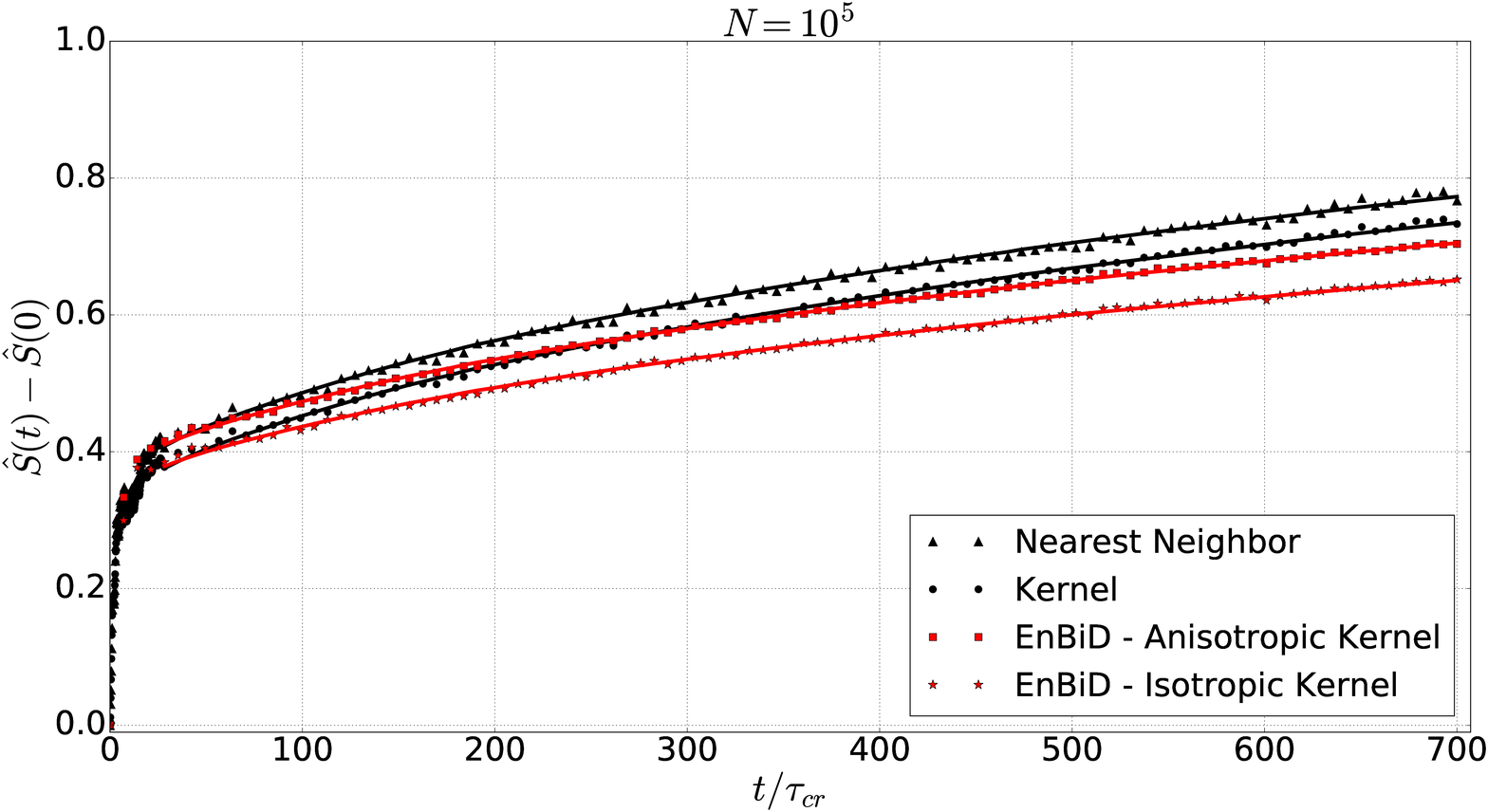}
  \caption{Same as Fig. \ref{fig:neig_vs_kern_vs_EnBiD_100k_short},
    but now showing the long-term evolution of the entropy. We see
    that all estimators provide the same behavior for the entropy,
    which increases almost linearly in the $2$-body relaxation
    time-scale. Continuous curves represent the orbit-averaged
    Fokker-Planck fit, estimated with Agama library.}
  \label{fig:neig_vs_kern_vs_EnBiD_100k_long}
\end{figure*}

\begin{figure*}[ht!]
  \epsscale{0.85}
  \plotone{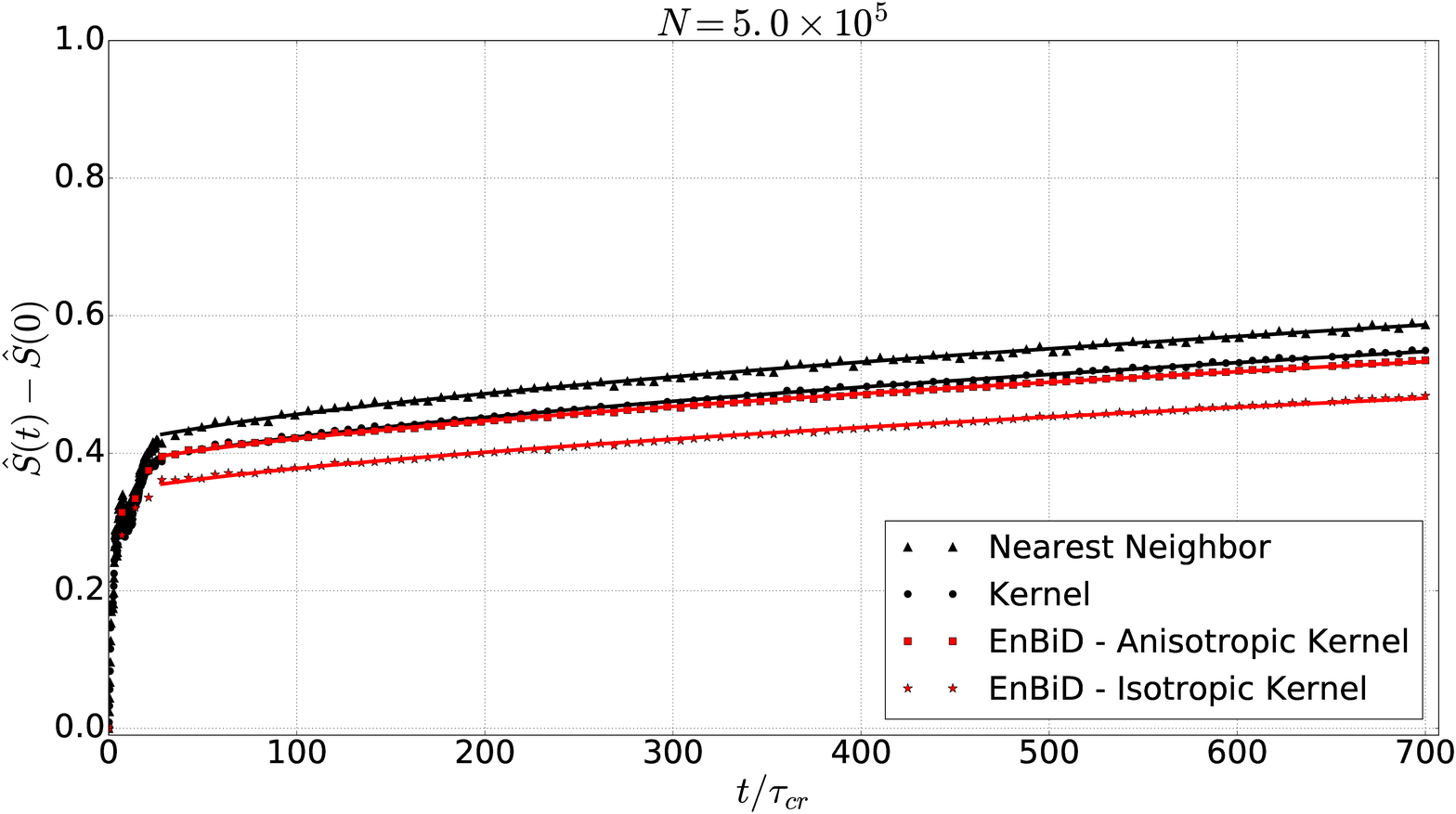}
  \caption{Same as Fig. \ref{fig:neig_vs_kern_vs_EnBiD_100k_long}, but
    now for $N=5\times 10^5$ particles. The agreement between Kernel
    method and EnBiD with anisotropic smoothing kernel is almost
    perfect.}
  \label{fig:neig_vs_kern_vs_EnBiD_500k_long}
\end{figure*}

Here we compare the long-term entropy evolution obtained with the
three different distribution function estimator methods, as explained
in
\S~\ref{sec:neig_vs_kernel}. Figs. \ref{fig:neig_vs_kern_vs_EnBiD_100k_long}
and \ref{fig:neig_vs_kern_vs_EnBiD_500k_long} show the results for
$N=10^5$ and $N=5\times 10^5$, respectively. In these figures, we can
see that the long-term evolution obtained with all methods are very
similar already for $N=10^5$ particles, with only a slightly smaller
slope for the EnBiD estimate. For $N=5\times 10^5$,
Fig.\ref{fig:neig_vs_kern_vs_EnBiD_500k_long}, the agreement of the
methods is even better, mainly the Kernel and EnBiD with anisotropic
smoothing kernel methods. This indicates again that the estimates,
which involve different techniques from each other, are capturing real
physical effects, and are not produced by numerical features.

\subsection{Dependence on the number of particles}
\label{sec:N_dependence_long}
\begin{figure*}[ht!]
  \epsscale{0.85}
  \plotone{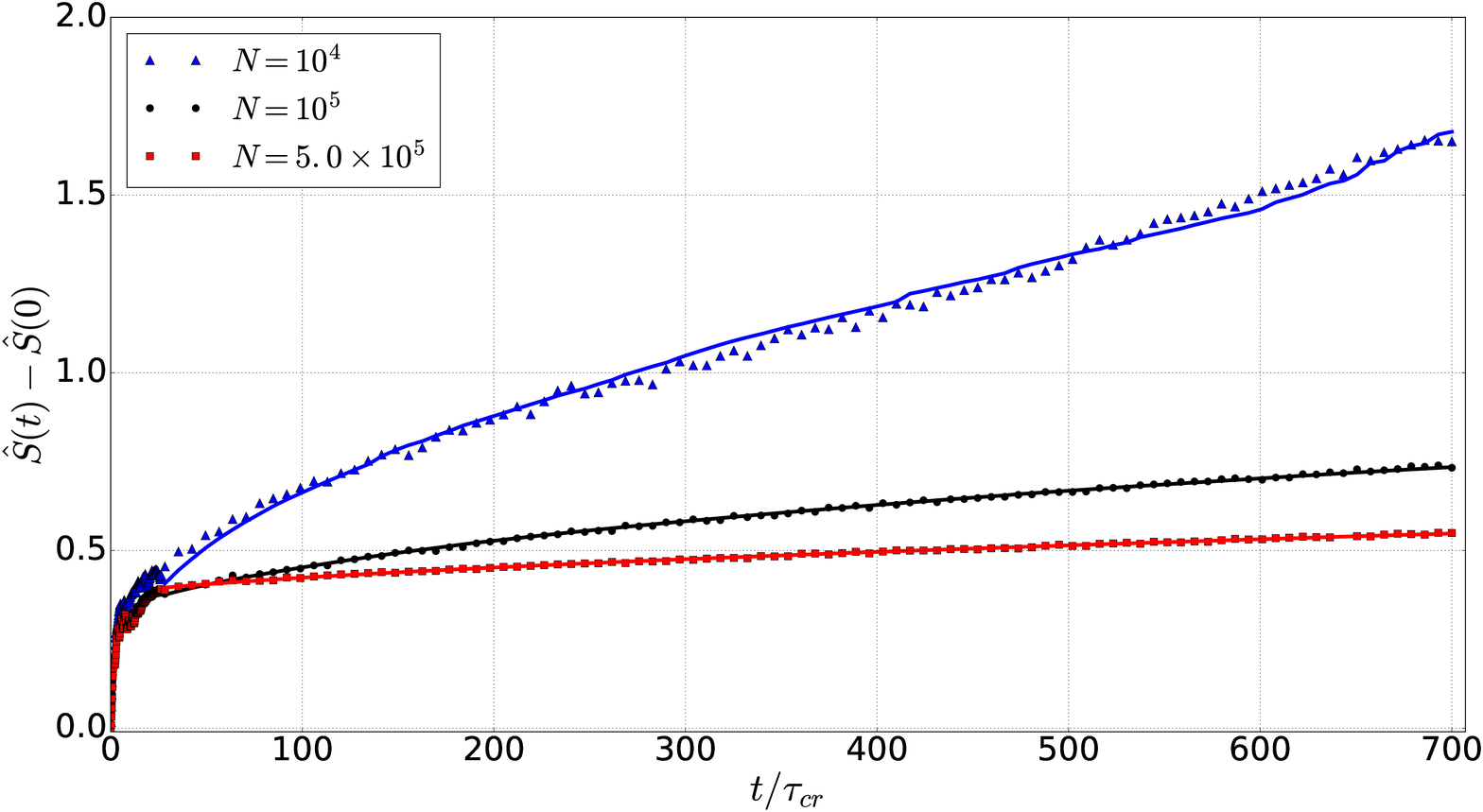}
  \caption{Entropy estimation for $N=10^4$ (blue triangles), $N=10^5$
    (black dots) and $N=5\times 10^5$ (red squares). Similarly to
    Fig. \ref{fig:fit_100k} but now comparing different number of
    particles. The entropy increase during violent relaxation is
    similarly high for the three data sets. However, for the long-term
    evolution, during which we expect the $2$-body relaxation to be
    significant, the slope of $S(t)$ is larger for a smaller number of
    particles $N$, in agreement with the theoretical expectation that
    $2$-body relaxation rate is larger for smaller $N$ -- see
    Eq.~\eqref{eq:tau_col}.}
  \label{fig:fit_10k_100k_500k}
\end{figure*}

In Fig. \ref{fig:fit_10k_100k_500k} we compare the long-term entropy
evolution for three different numbers of particles: $N=10^4$ (blue
triangles), $N=10^5$ (black dots) and $N=5\times 10^5$ (red squares),
as well as the fit of the orbit-averaged Fokker-Planck equation for
$t/\tau_{cr}\gtrsim 26.5$. We see that the prediction for $N=10^4$ is
a little bit noisy. In principle, this can be attributed to two
causes: shot noise due to a relatively small number of particles or
non-validity of some hypotheses on which our Fokker-Planck model is
based. In \S\ref{sec:eps_long} we present arguments in favour of this
second option. More specifically, it seems that when $N$ is small we
still have a substantial number of almost close encounters
($b\gtrsim b_0$), producing scattering angles $\lesssim 90^\circ$,
which can be enough to violate the weak coupling assumption on the
basis of the Fokker-Planck treatment -- see \cite{Binney_2008},
sec. 7.4.4.. However, again we see that in general the Fokker-Planck
equation is able to explain the long-term entropy evolution. Taking
into account the estimate of the $2$-body relaxation time-scale,
Eq.~\eqref{eq:tau_col}, the fact that the slope of $S(t)$ is larger
for decreasing number of particles shows that the $2$-body relaxation
is in fact more effective for a smaller number of particles, in
agreement with the theoretical expectation: in the limit
$N\rightarrow\infty$ we would have a collisionless system and thus the
slope of $S(t)$ predicted by the Fokker-Planck equation would be zero.

\begin{figure*}[ht!]
  \epsscale{0.85}
  \plotone{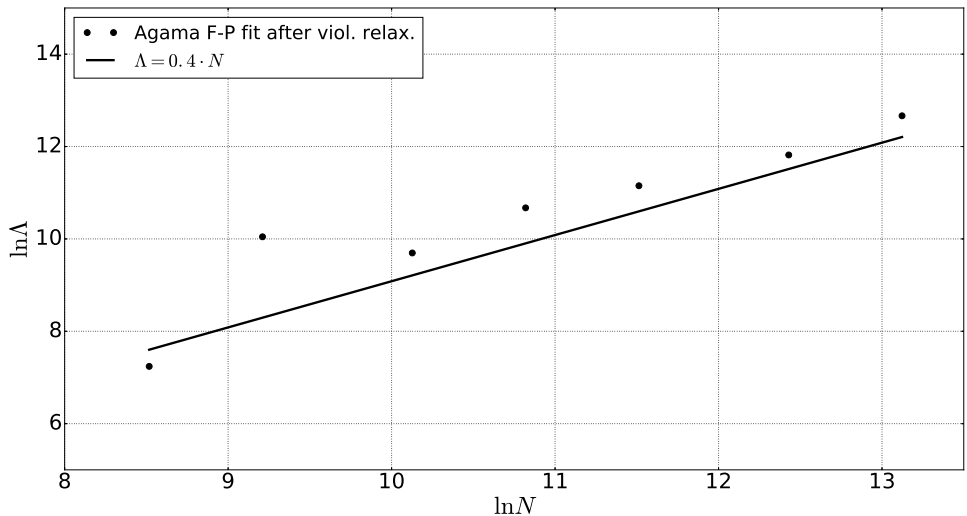}
  \caption{Coulomb logarithm obtained fitting Agama modelling of
    orbit-averaged Fokker-Planck equation to the long-term entropy
    evolution for different numbers of particles (black dots). We can
    see a reasonable agreement with the theoretical expression
    $\ln\Lambda = 0.4 N$ shown as the black straight line.}
  \label{fig:log_Lambda_vs_N}
\end{figure*}

With this long-term entropy evolution for different numbers of
particles, we can also study the $N$-dependence of the $2$-body
relaxation time-scale, Eq.~\eqref{eq:tau_col}. More specifically, we
study the $N$-dependence of the Coulomb logarithm $\ln \Lambda$
obtained with the fit to this long-term evolution. This is represented
by the black dots in Fig.~\ref{fig:log_Lambda_vs_N}. From the
theoretical point of view, the Coulomb logarithm is given by
${\ln \Lambda=\ln b_{max}/b_0}$, where $b_{max}$ is usually assumed to
be of order of the system's size (although some authors support the
idea that $b_{max}$ should be equal to the mean interparticle
distance, i.e.  $b_{max}\propto N^{-1/3}$ -- see
\cite{1942ApJ....95..489C,Kandrup_1980, 1982ApJ...253..512F,
  1992ApJ...398..519S, 1994ApJ...427..676F, 1998A&A...330.1180T}). In
what follows, we set ${b_{max}= R_{hm}}$, the half-mass radius
\citep[see][]{1987degc.book.....S}. The impact parameter associated to
a $90^\circ$ scatter is given by:
\begin{equation}
  \label{eq:b_0_def}
  b_0= 2\frac{Gm}{V^2},
\end{equation}
where $V$ is the relative velocity between the test and field
particles (here again assumed to have equal mass $m$).

The virial radius $R_{vir}$ is defined by (see Appendix
\ref{sec:units})
\begin{equation*}
  \langle v^2\rangle = \frac{GM}{2R_{vir}}.
\end{equation*}
Substituting in Eq.~\eqref{eq:b_0_def} we obtain
\begin{equation}
  \label{eq:11}
  \frac{R_{vir}}{b_0}=\frac{N}{2}\frac{V^2}{2\langle v^2\rangle}.
\end{equation}
Assuming a Plummer density profile, the half-mass radius is given by
$R_{hm}\approx 0.8 R_{vir}$ and with
$\langle V^2\rangle =2\langle v^2 \rangle$ and assuming the system is
virialized (see Appendix \ref{sec:units}), we finally have
\begin{equation}
\label{eq:Lambda}
\Lambda =\frac{R_{hm}}{b_0} \approx 0.4N.
\end{equation}


In Fig. \ref{fig:log_Lambda_vs_N}, this relation is shown as the black
straight line. We see that despite some discrepancy for small $N$, the
data points show a reasonable agreement with Eq.~\eqref{eq:Lambda}.


\subsection{Comparison of different N-body codes and the role of the
  softening length}
\label{sec:eps_long}
\begin{figure*}[ht!]
  \epsscale{0.85}
  \plotone{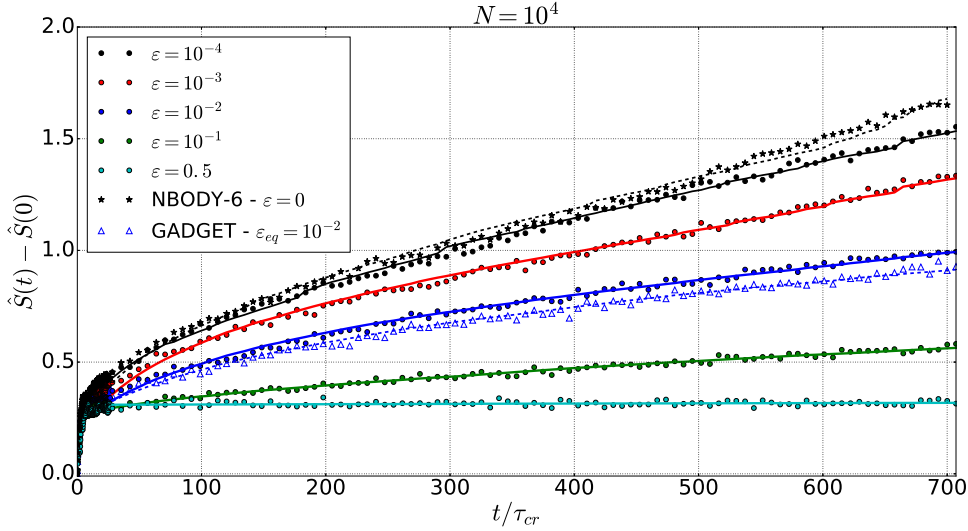}
  \caption{Long-term evolution of entropy estimation for $N=10^4$
    particles obtained with NBODY-2 for different values of the
    softening length $\varepsilon$. As expected, suppression of
    collisional relaxation is larger for larger $\varepsilon$,
    saturating just at scales larger than $\bar{d}$. This favours
    $b_{max}=R$ -- see \S~\ref{sec:Fokker_Planck}. For
    $\varepsilon =10^{-4}\approx R/N$, the entropy evolution is
    identical to that without softening length (stars and black dashed
    line) obtained with NBODY-6, confirming that close encounters,
    i.e. those involving impact parameters $b\lesssim b_0\approx R/N$,
    do not contribute significantly to relaxation. Also shown the
    entropy obtained with GADGET-2 (blue open triangles) for a
    softening length $h=2.8\times 10^{-2}$, i.e. with a
    Plummer-equivalent softening length $\varepsilon_{eq}=10^{-2}$,
    whose evolution is very similar to that of NBODY-2 with the same
    $\varepsilon$ (blue dots).}
  \label{fig:10k_eps_long}
\end{figure*}

Similarly to \S~\ref{sec:eps}, we now compare the long-term entropy
evolution obtained with the three different N-body simulation codes
used (NBODY-6, NBODY-2 and GADGET-2) with different values of the
softening length, when it applies. Fig.~\ref{fig:10k_eps_long} shows
the long-term entropy evolution and the fit of the Fokker-Planck model
for different values of $\varepsilon$ obtained with NBODY-2, now
fixing $N=10^4$. The stars and the black dashed line represent the
case obtained with NBODY-6, i.e. without softening length. Firstly, we
can see that for ${\varepsilon = 10^{-4} \approx R/N}\approx b_0$ (see
Eq.~\eqref{eq:Lambda}), the entropy evolution is very similar to that
without softening length, i.e. encounters with impact parameter
$b\lesssim b_0$, i.e. close encounters, do not seem to contribute
significantly to the relaxation, confirming earlier results from
\cite{1982ApJ...253..512F}.

On the other hand, we clearly see that larger values of the softening
length increasingly suppress the $2$-body relaxation, decreasing the
slope of $\hat{S}(t)$, even for
$\varepsilon > \bar{d}=R/N^{1/3} \approx 0.05$. This indicates that
scatterings involving distances $>\bar{d}$ seem to be important for
the $2$-body relaxation, an indication in favor of
${b_{max}\approx R}$, in contrast to $b_{max}=\bar{d}$ as predicted by
some authors \citep[see][]{1994ApJ...427..676F}, although more
detailed analysis, with larger $N$, would be necessary.


In Fig.~\ref{fig:10k_eps_long} the blue open triangles represent the
entropy evolution obtained from GADGET-2 with $N=10^4$, the same
initial conditions as before and with a softening length
$h=2.8\times 10^{-2}$, i.e. with a Plummer-equivalent softening length
$\varepsilon_{eq}=10^{-2}$ -- see \S \ref{sec:simulations}. Also shown
is the fit of the Fokker-Planck model (dashed blue line), from which
we see that the 2-body relaxation is slightly more suppressed with
GADGET-2 than with NBODY-2 with equivalent softening lengths, but
their overall evolutions are very similar. These results confirm that,
even though codes such as GADGET-2 are frequently called
collisionless, they can still present significant collisional
relaxation in the long-term evolution. Furthermore, this collisional
relaxation is similar to that obtained with other techniques and
equivalent softening lengths, as already demonstrated by
\cite{1990ApJ...349..562H, 2015MNRAS.453.2919S}.

Regarding the discussion raised in \S\ref{sec:N_dependence_long} about
the noisier appearance of the Fokker-Planck prediction for $N=10^4$ in
comparison to larger $N$, it is interesting to see that for increasing
$\varepsilon$ the prediction rapidly becomes smoother, even though the
number of particles is fixed. This seems to indicate that the noise is
not directly due to Poisson fluctuations of a small number of
particles. Instead, since the introduction of a softening length
suppresses encounters producing large scattering angles (even though
not strictly close encounters, for which $b < b_0$, as shown above),
the noise seems to be associated with the presence of a substantial
number of such almost close encounters ($b\gtrsim b_0$, producing
scattering angles $\lesssim 90^\circ$), what is expected for small
$N$.

\begin{figure*}[ht!]
  \epsscale{0.85}
  \plotone{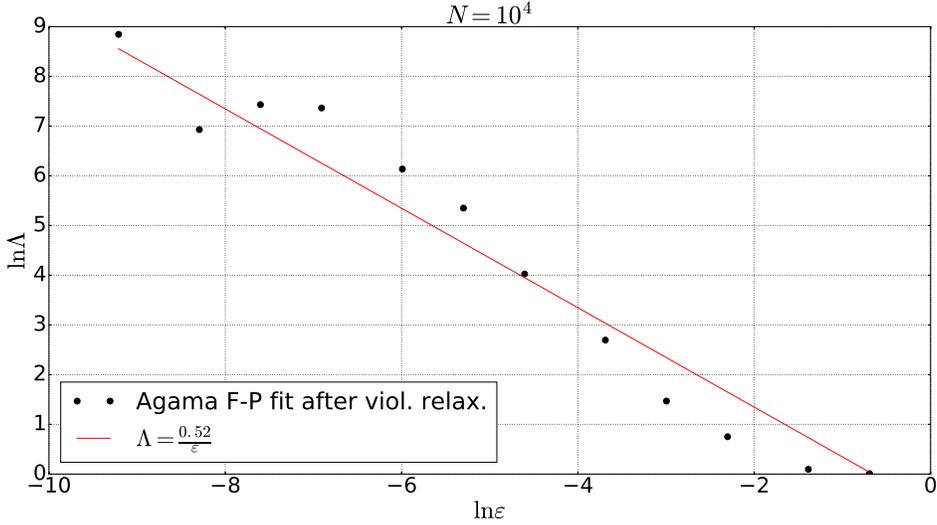}
  \caption{Coulomb logarithm from the fit of orbit-averaged
    Fokker-Planck equation to the entropy evolution obtained with
    NBODY-2 for $N=10^4$ as a function of the softening length
    $\varepsilon$.}
  \label{fig:log_Lambda_vs_eps_10k}
\end{figure*}

In Fig. \ref{fig:log_Lambda_vs_eps_10k} we show the Coulomb logarithm
obtained fitting the Fokker-Planck equation to the entropy evolution
for different $\varepsilon$ values (dots). The effect of suppression
on 2-body relaxation for increasing softening length $\varepsilon$ can
be parametrized in $\ln\Lambda$ substituting $b_0$ in
Eq.~\eqref{eq:Lambda} by an effective impact parameter
${b_{eff} = b_{min} + \varepsilon}$, where $b_{min}$ is expected to be
of order of $b_0\approx R/N$, as suggested by
\cite{2003MNRAS.344...22S}. The red curve shows the fit of
$\ln\Lambda = \ln\left[ b_{max}/(b_{min} + \varepsilon)\right]$, where
$b_{max}$ and $b_{min}$ are free parameters, for which we get
$b_{max}=0.52$ and $b_{min} \approx 0.0$. The value obtained for
$b_{max}$ agrees with the assumption made in
\S~\ref{sec:N_dependence_long}, where we used $b_{max}=R_{hm}$. On the
other hand, the value for $b_{min}$ is significantly smaller than
expected, being compatible with zero. This seems to be associated to
the fact that the first data point (smallest $\varepsilon$) is
significantly higher than the expected trend of
$\Lambda \approx const$ in this region.

\section{Conclusions}
\label{sec:conclusions}
As discussed in \S \ref{sec:introduction}, the validity of the
Vlasov-Poisson equation implies the conservation of quantities defined
by Eq. \eqref{eq:1}, for any functional $s[f]$ -- see
\cite{Tremaine_Henon_Lynden_Bell_1986}. Conversely, non-conservation
of any such quantity, like the Shannon entropy, implies non-validity
of Vlasov-Poisson equation. In this paper, through the use of $N$-body
simulations, we estimate the entropy at each time step, following its
time evolution in order to test the validity of the Vlasov-Poisson
equation for a collisionless self-gravitating system during violent
relaxation.

Our results show a clear separation between the two relevant time
scales: the dynamical time scale and the $2$-body relaxation time
scale. During the early stages, in the dynamical time scale, the
entropy has a significant increase accompanied by damping oscillations
-- see Fig. \ref{fig:100k_short}. This is the time scale in which the
violent relaxation process is known to operate, the relaxation being
associated to the typical particle interaction with the time-changing
collective gravitational potential. This is the main result of this
work, indicating that the Vlasov-Poisson equation does not seem to be
valid during violent relaxation. This conclusion provides a natural
solution to the so-called ``fundamental paradox of stellar dynamics''
\citep[][]{1965dss..book.....O, 2006AAT...25..123O}. Several tests,
changing the $N$-body simulation setups and estimators, as discussed
in the preceding sections and commented below, lead to the same
conclusion.

Instead of representing an exceptional situation, the entropy
oscillations observed reinforce the reliability of our results, since
they are typical of systems with considerable potential energy, as
pointed out by
\cite{1966Phy....32.1376P,1971PhRvA...4..747J,1997JSP....89..735R}. Moreover,
\cite{1971PhRvA...4..747J} argues that the fact the ``H'' function is
not monotonic can be associated to the non-validity of the hypothesis
of molecular chaos, which is of fundamental importance in the
derivation of the Vlasov-Poisson equation through the BBGKY hierarchy,
a possibility also suggested by \cite{Beraldo_Lima_Sodre_Perez_2014}.

Studying the impact of different initial conditions, the general
conclusion is that entropy increase is higher for initial conditions
farther from equilibrium. In particular, starting the simulation with
the self-consistent Plummer model, which is a steady state, we see
practically no entropy increase. This is in agreement with the general
idea behind the statistical interpretation of the second law of
Thermodynamics: that the systems evolve to the most probable state and
that the entropy increase is due to the particular choice of a initial
state which is very unlikely in comparison to the equilibrium
state. The fact that we observe a negligible entropy increase in the
case of initial Plummer model also weakens the possibility that the
entropy increase observed in the other runs is being produced by some
artificial, numerical relaxation.

Regarding the number of particles, we see in
Fig.~\ref{fig:10k_100k_1m_short} that the early evolution of the
entropy is qualitatively similar for $N=10^4$, $N=10^5$ and
$N=10^6$. The more irregular appearance of the data points for
$N=10^4$ can be attributed to shot noise due to the relatively small
number of particles, while for $N=10^5$ and $N=10^6$ the curves are
pretty smooth.

We note that oscillation amplitudes are larger for $N=10^6$ in
comparison to $N=10^5$. This can be interpreted as follows: the
$2$-body relaxation, although globally negligible for entropy increase
already for $N=10^5$, can still act to destroy the coherent
oscillatory pattern, smoothing out the curve, while for $N=10^6$ the
$2$-body relaxation has even less effect, allowing the presence of
coherent oscillations for a longer time period. This also reinforces
the idea that convergence to significantly large $N$, i.e. to the
collisionless regime, has been achieved for the time-scale of violent
relaxation.

We also tested three different methods for estimating the distribution
function: the Kernel method, Nearest Neighbor method and EnBiD
\citep{2006MNRAS.373.1293S}. For $N=10^5$,
Fig.~\ref{fig:neig_vs_kern_vs_EnBiD_100k_short} shows that Kernel and
Nearest Neighbor have a general agreement with each other, but EnBiD
shows some differences in respect to those methods, mainly a larger
initial entropy production. Nevertheless, the general qualitative
behaviour is similar for all these methods: a high entropy increase
followed by damping oscillations in the dynamical time-scale. For
$N=10^6$, Fig.~\ref{fig:neig_vs_kern_vs_EnBiD_1m_short} shows a much
better agreement of all the methods, mainly between Kernel method and
EnBiD with anisotropic kernel smoothing. This agreement suggests that
the entropy evolution observed is not being produced by numerical
artifacts but represents a good estimator for the true entropy
evolution.

We also observe a good agreement for the entropy evolution obtained
with the three different N-body simulation codes NBODY-6, NBODY-2 and
GADGET-2 despite the fact the these codes involve different
integration techniques. This reinforces the idea that the entropy
evolution observed is not a numerical effect, but a real physical
effect. Therefore, the results obtained from NBODY-2 with different
values of the softening length $\varepsilon$ suggest that violent
relaxation involves spatial scales around the mean neighboring
particle distance, in agreement with the mathematical results
summarized in Appendix~\ref{sec:math_vlasov}.

On the other hand, in the long-term, i.e. in the collisional
relaxation time scale the entropy increases almost linearly. In order
to verify if this entropy increase can be produced by $2$-body
relaxation, we estimate the contribution of the orbit-averaged
Fokker-Planck model to the entropy increase, Eqs.~\eqref{eq:est_S_dot}
and~\eqref{eq:FP}. Fig.~\ref{fig:fit_100k} shows that this is the case
indeed: if we restrict the fit to points after the early stages, say
for $t/\tau_{cr}\gtrsim 26.5$, the Fokker-Planck model can reproduce
quite accurately the almost linear entropy increase. However, not
surprisingly, the orbit-averaged Fokker-Planck model cannot fit the
early entropy evolution, during violent relaxation, where the
potential is time-changing and the assumption $f=f(E)$ does not apply.

An interesting observation is that while the effectiveness of violent
relaxation, i.e. the slope of $S(t)$, depends crucially on the initial
conditions, the $2$-body relaxation seem to depend only on the number
of particles $N$.

The variation of the softening length $\varepsilon$ in the long-term
simulations generated with NBODY-2 also allowed us to study the
characteristic scales for $2$-body relaxation. Our results seem to
confirm that this relaxation process involve scales within the
interval ${R/N\leq b\leq R}$.

In conclusion, our results show that during violent relaxation there
is a significant entropy increase despite the prediction of the
Vlasov-Poisson equation of entropy conservation. Under the assumption
that the convergence for significantly large $N$ has been achieved
(which seems to be the case), this early entropy increase cannot be
attributed to any $2$-body relaxation process, and must be associated
to a collective effect, which is the original idea behind violent
relaxation.

These results indicate that the Vlasov-Poisson equation is not valid
for collisionless self-gravitating systems during violent relaxation,
resurrecting the arrow of time in the collapse of these systems.

Finally, the fact that the early regime with fast entropy changes
extends up to $t\approx 25 \tau_{cr}$, i.e. times substantially larger
than $\tau_{cr}$ but still substantially smaller than $\tau_{col}$,
can possibly support the collective relaxation time scale
$\tau \propto N^{1/3}\tau_{cr}$ predicted by
\cite{1986A&A...160..203G}. However, due to the reasonably weak
$N$-dependence, a more rigorous test would require significantly
larger numbers of particles.

\section{Final remarks}
\label{sec:final_remarks}

The second law of Thermodynamics refers to the macroscopic evolution
of any physical system, and does not depend on its on a description in
terms of its microscopic constituents. Therefore, whenever a
\emph{macroscopically} irreversible evolution takes place, entropy is
expected to increase. All results discussed in this work are in
accordance with this idea.

On the other hand, when trying to explain the entropy evolution by
means of a transport equation, it is important to keep in mind that
even though the equation refers to the coordinates of one particle, it
is not an equation of motion for some specific, randomly chosen,
particle or fluid element. Instead, it statistically expresses the
evolution of the system as a whole, referring to the coordinates of a
statistical entity, the typical (or test) particle. As a result, it
captures the influence of collective effects, which cannot be achieved
using the equations of motion for a single particle. In this respect,
any ``transport equation'' whose derivation is based only on
mechanical considerations cannot describe the evolution of the system
as a whole, but merely reassert the Hamiltonian evolution of each
particle individually.

We should also mention that the right hand side of a transport
equation is associated to any process relaxing the system, which only
in the case of a molecular gas is necessarily realized by collisions,
i.e. $2$-body interactions. This point is particularly important for a
system with long-range interactions, which can relax even if it is
collisionless\footnote{This is the reason why we prefer not to call
  Eq.~\eqref{eq:vlasov} the ``collisionless Boltzmann equation'' as
  recommended by \cite{1982A&A...114..211H}.}. In fact, taking the
presence of chaotic motions as a diagnostic of relaxation, it is
interesting to remember that $N$-body self-gravitating systems can
exhibit large (and increasing as $\ln N$) rates of growth of small
perturbations, even for large $N$ -- see
\cite{2002ApJ...580..606H}. In other words: these systems seem to be
more chaotic for larger $N$, i.e. when they approach the collisionless
regime -- see also \cite{2003ApJ...585..244K}.

The original statistical meaning of the distribution function seems to
be frequently neglected, and in our opinion, the standard view of the
necessity of a coarse-grain interpretation for the macroscopic
evolution of the system serves to introduce this statistical
meaning. This coarse-grain interpretation also seems to be reminiscent
of discussions regarding the difference between Gibbs's and
Boltzmann's definitions of entropy
\citep{1993PhT....46i..32L,1993PhyA..194....1L,1965AmJPh..33..391J,
  2001LNP...574...39G}, which is in the heart of the opposition
between the microscopic and macroscopic descriptions of the evolution
of any system, not being restricted to the gravitational $N$-body
problem. While Gibb's entropy makes reference to the N-particle
distribution function $f^{(N)}$, whose content is exclusively
mechanical and is subject to Liouville's equation $df^{(N)}/dt=0$,
thus being conserved, Boltzmann's entropy is defined in terms of the
one-particle distribution function $f$, which has a statistical
content and evolves in time. According to \cite{1965AmJPh..33..391J},
``{\it since the Gibbs H is dynamically constant, one has resorted to
  some kind of coarse-graining operation, resulting in a new quantity
  $\bar{H}$, which tends to decrease (...). Mathematically, the
  decrease in $\bar{H}$ is due only to the artificial coarse-graining
  operation and it cannot, therefore have any physical significance
  (...). The difference between H and $\bar{H}$ is characteristic, not
  of the macroscopic state, but of the particular way in which we
  choose to coarse-grain. Any really satisfactory demonstration of the
  second law must therefore be based on a different approach than
  coarse-graining}''. As we can see, the standard coarse-grain
interpretation is only necessary (though subject to criticism) when we
adopt Gibbs's entropy definition, i.e. in terms of $f^{(N)}$. However,
if we adopt Boltzmann's definition in terms of $f$, as done in this
work, any possible coarse-grain interpretation is intrinsically there
and any extra coarse graining is not necessary once we recognize the
statistical content of the distribution function and abandon the
assumption of validity of the Vlasov-Poisson equation during violent
relaxation.

\appendix
\section{Summary of mathematical results on the Vlasov-Poisson
  equation}
\label{sec:math_vlasov} 

For the Vlasov-Poisson initial value problem itself, there are various
results \citep{Pfaff,Sch,LiPerth,Horst} ensuring global existence and
uniqueness of weak and strong solutions under fairly general
conditions on the initial configuration. The first mathematically
rigorous derivations of the Vlasov equation\footnote{For a general
  overview of the topic, we refer the reader to the monograph
  \cite{Spohn}.} from a many-body problem can be found, e.g., in
\cite{NeunWick,BraunHepp,Dobr,Neun}. Rather than the Vlasov-Poisson
equation, \cite{NeunWick,BraunHepp,Dobr,Neun,Spohn} consider models
with \emph{continuous and bounded} forces. In particular, these forces
are not diverging at small inter-particle distances $D$, in contrast
to gravitational (Coulomb) forces.  In the last few years progress has
been made in treating mean field limits for forces which are singular
at small distances up to but not including the Coulomb
case. \cite{HauJa} discusses forces diverging as $\sim D^{-\alpha }$
with $\alpha <d-1$ in $d\geq 3$ dimensions. More precisely, in the
case $1<\alpha <d-1$, a softening length of order $N^{-\frac{1}{2d}}$
is assumed in the derivation of the Vlasov equation from the $N$-body
problem \citep{HauJa}. If $\alpha <1$ no positive softening length is
needed in the proof. \cite{Kiess} managed to prove a result including
the Coulomb singularity under the assumption of an (uniform in $N$) a
priori bound on the microscopic forces. However, whether it is
satisfied for generic initial data or not, remains an open problem
\citep{Kiess}. \cite{BoePick} improved the result of \cite{HauJa} in
the sense that the softening length used is of order
$N^{-\frac{1}{d}}$, but still $\alpha$ has to be strictly smaller than
$d-1$ (and the Coulomb case is again not included). Recently,
\cite{La,LaPick} extended the method of \cite{BoePick} to include the
Coulomb singularity, in 3 dimensions, aiming at a microscopic
derivation of the Vlasov-Poisson dynamics. As in \cite{HauJa}, a
strictly positive softening length is needed, at fixed $N$. It can be
chosen, as shown in \cite{LaPick}, of order $N^{-\beta }$ with
$\beta <\frac{1}{3}$.

These analyses suggest that inter-particle interactions taking place
up to distances $d_{0}$ that are large compared to the mean
neighboring particle distance $\bar{d}$, but small compared to the
size of the whole system (i.e., scattering processes at impact
parameters $b$ with $b\lesssim d_{0}$ and $\bar{d}\ll d_{0}\ll 1$),
could prevent Vlasov-Poisson equation from being the effective
macroscopic equation governing the evolution of collisionless
self-gravitating systems. Note also that the long-range character of
potentials is harmless in the mathematical derivation of Vlasov
equations for $N$-body systems, as soon as their gradients (forces)
are uniformly bounded at large distances. From this, we would not
expect the emergence of the arrow of time in collisionless
self-gravitating systems to be a consequence of scattering processes
at large impact parameters.

\section{Some useful quantities and the H\'enon units}
\label{sec:units}
The virial ratio is defined as
\begin{equation}
  \label{eq:3}
  Q = -\frac{T}{W},
\end{equation}
where $T$ is total kinetic energy and $W$ is total potential energy. A
convenient scale length is the virial radius $R_{vir}$ defined by
\begin{equation}
  \label{eq:R_vir_def}
  W = -\frac{GM^2}{2R_{vir}},
\end{equation}
where $G$ is the gravitational constant and $M$ is system's total
mass. The rms velocity is defined by
\begin{equation}
  \label{eq:5}
  \langle v^2\rangle = \frac{2T}{M} = \frac{2(E-W)}{M},
\end{equation}
where $E$ is system's total energy. The mean crossing time $\tau_{cr}$
is defined by
\begin{equation}
  \label{eq:6}
  \tau_{cr} = \frac{2R_{vir}}{\sqrt{\langle v^2\rangle}}.
\end{equation}

The H\'enon units \citep[see][]{Henon_1964}, also called N-body units,
are defined making $G=1$, $M=1$ and $E=-1/4$. Substituting in the
above expressions, we obtain
\begin{equation}
  \label{eq:8}
  R_{vir} = 2(1-Q),
\end{equation}
\begin{equation}
  \label{eq:9}
  \sqrt{\langle v^2\rangle} = \sqrt{\frac{Q}{2(1-Q)}}
\end{equation}
and thus 
\begin{equation}
  \label{eq:10}
  \tau_{cr}=4\sqrt{2}\sqrt{\frac{(1-Q)^3}{Q}}.
\end{equation}
In virial equilibrium, the total kinetic and
potential energies respect
\begin{equation}
  \label{eq:2}
  W = -2T,
\end{equation}
i.e. $Q=1/2$. Thus, in virial equilibrium, we have $R_{vir}=1$,
$\langle v^2 \rangle = 1/2$ and finally $\tau_{cr}=2\sqrt{2}$.

Regarding the definition of the Plummer model, its density profile is
given by
\begin{equation}
  \label{eq:7}
  \rho(r) = \frac{3M}{4\pi a^3}\frac{1}{\left[1 + \left(r/a\right)^2\right]^{5/2}},
\end{equation}
where $M$ is total mass and $a$ is a scale parameter. Thus, the radius
containing half of the total mass is $R_{hm}\approx 1.3 a$, and the
total potential energy is $W = -3\pi GM^2/(32a)$. Substituting the
$R_{vir}$ definition, Eq.~\eqref{eq:R_vir_def}, we have
$R_{vir}=16a/(3\pi)$. Thus, in the Plummer model
$R_{hm}\approx 0.8R_{vir}$, and if the system is virialized,
$R_{hm}\approx 0.8$ in H\'enon units. 

Regarding the impact parameter associated to a $90^\circ$ scattering
angle,
\begin{equation}
  \label{eq:15}
  b_0 = \frac{G(m + m_f)}{V^2},
\end{equation}
the relative velocity between the test and field particle is
$\vec{V} = \vec{v} - \vec{v}_f$. Thus,
$V^2 = v^2 + v_f^2 - 2\vec{v}\cdot\vec{v}_f$. Averaging over all
particles of the system, we have
$\langle V^2 \rangle = \langle v^2\rangle + \langle v_f^2\rangle$. And
if $m = m_f$, we have $\langle V^2\rangle= 2\langle v^2\rangle$. Thus,
for a virialized system composed of equal mass particles, we have
\begin{equation}
  \label{eq:16}
  \Lambda = \frac{R_{hm}}{b_0} \approx 0.4 N.
\end{equation}

\section*{Acknowledgements}
We are very grateful to S. Aarseth for all the help with the codes
NBODY-6 and NBODY-2, to Monica Valluri for a critical reading and for
suggesting the use of EnBiD method. We are thankful also to Eugene
Vasiliev for the help with Agama code and discussions regarding the
orbit-averaged Fokker-Planck modelling. We also thank Jean-Bernard Bru
and Roberto Venegeroles for discussions. This work has made use of the
computing facilities of the Laboratory of Astroinformatics (IAG/USP,
NAT/Unicsul), whose purchase was made possible by the Brazilian agency
FAPESP (grant 2009/54006-4) and the INCT-A. We are specially grateful
to Carlos Paladini, without whom this work would not have been
possible. LBeS and ELDP are supported by FAPESP. LSJ acknowledges
support by FAPESP (2012/00800-4) and CNPq. ML is also partially
supported by FAPESP and CNPq. WdeSP is partially supported by FAPESP,
CNPq and the Spanish Ministry of Economy and Competitiveness (grant
SEV-2013-0323).

\bibliography{/home/leandro/references_leandro}

\end{document}